\documentclass[12pt]{article}

\usepackage{newtxtext,newtxmath}
\usepackage{braket}
\usepackage{graphicx}

\usepackage[letterpaper,margin=1in]{geometry}

\renewenvironment{abstract}
	{\quotation}
	{\endquotation}

\date{}

\makeatletter
\renewcommand{\fnum@figure}{\textbf{Figure \thefigure}}
\renewcommand{\fnum@table}{\textbf{Table \thetable}}
\makeatother

\usepackage{scicite}

\usepackage{url}

\def\scititle{
	Polaritonic Bloch's theorem beyond the Long-Wavelength Approximation
}
\title{\bfseries \boldmath \scititle}

\author{
	Giovanna Bruno$^{1\ast}$,
	Rosario R. Riso$^{2}$,
	Henrik Koch$^{2}$ and Enrico Ronca$^{1\ast}$
    \and
	\small$^{1}$Dipartimento di Chimica, Biologia e Biotecnologie, Università degli Studi di Perugia, Perugia \& 06123, Italy.\and
	\small$^{2}$Department of Chemistry, Norwegian University of Science and Technology, Trondheim \& 7491, Norway.\and
	\small$^\ast$Corresponding author. Email: giovanna.bruno@unipg.it;
           enrico.ronca@unipg.it}

\begin{document} 

\maketitle

\begin{abstract} \bfseries \boldmath
Cavity quantum electrodynamics provides a powerful tool to manipulate material properties, yet it remains a matter of debate whether and how quantized fields affect the periodicity of crystals. Here, we extend Bloch’s theorem to crystals under strong light–matter coupling, revealing that polariton quasiparticles preserve lattice periodicity. We introduce a general framework to incorporate multimode cavity fields in a simple and tractable way, showing that additional modes contribute small energy corrections noticeable only at low frequencies. Within the single-photon approximation, these contributions reduce to a spatially uniform effective field in the crystal plane, providing a formal justification for the single-mode and long-wavelength approximations commonly used in molecular polaritonics. Together, these results establish a rigorous framework for describing polaritonic states in crystalline solids.
\end{abstract}

\noindent
In the last decade, strong light-matter coupling has emerged as a new frontier in materials science, offering unprecedented opportunities to engineer functionalities beyond conventional design. This regime occurs in optical cavities when coherent exchange between matter and the quantized field, sustained even by vacuum fluctuations, exceeds decoherence, giving rise to hybrid light–matter states (polaritons) which exhibit distinct features \cite{Haroche_Nobel_2013,Dovzhenko_2018,Hertzog_2019,Garcia_Vidal_2021}. This leads to phenomena as diverse as altered chemical reactivity \cite{Ebbesen_2012,Ebbesen_2016,Ebbesen_2019,Haugland_2021,Castagnola_2024}, modified optical absorption and emission \cite{Herrera_2017,DelPo_2020,Li_2021} and cavity-induced quantum phase transitions, including superconductivity\cite{Sentef_2018,Lu_2024,Kozin_2025}, quantum Hall conductance \cite{Ciuti_2021,Rubio_2022,Rokaj_2022,Appugliese_2022,Rokaj_2023}, metal-to-insulator transitions \cite{Jarc_2023,Jarc_2024,Fassioli_2024} and spin order modifications  \cite{Ashida_2020,Latini_2021,Vinas_2023}. These findings highlight the transformative potential of cavity quantum electrodynamics (cavity QED), especially for next generation quantum technologies \cite{Hubener_2024,Imamoglu_2025}. Therefore, increasing attention is being devoted to quantum materials, predominantly crystalline solids, whose rich phase diagrams and collective excitations provide a fertile ground to explore and exploit cavity-induced effects \cite{Liu_2015,Keimer_2017,Mazza_2019,Cava_2021,Schlawin_2022,Tan_2023,Wang_2025,Helmrich_2025}.
However, the control of these phenomena is still limited. While experimental progress is hindered by technical challenges and sensitivity to ambient parameters \cite{simpkins_2021}, the most critical issue lies in theory: without a comprehensive microscopic understanding of how strong coupling reshapes material properties, our possibility to predict, design and ultimately engineer these effects remains fundamentally restricted.\par 
So far, most of theoretical studies of crystalline materials in cavities have relied on phenomenological models \cite{Rabi_1936,Rabi_1937,Purcell_1946,Jaynes_Cummings_1963} which, while capturing essential physics, neglect key aspects such as electronic correlation, multimode effects and/or spatial inhomogeneity of the field \cite{Ciuti_2021,Rokaj_2023,Jarc_2023}. To provide more robust and transferable theories, some advanced \textit{ab-initio} approaches—including density functional theory (DFT) \cite{Tokatly_2013,Ruggenthaler_2014,Ruggenthaler_2015,Flick_2015,Fuck_2016}, Hartree–Fock \cite{Haugland_2020,Schnappinger_2023,Castagnola_2025}, and coupled-cluster \cite{Haugland_2020,Ruggenthaler_2023,Foley_2023,Castagnola_2023}—have been extended to the strong coupling regime, albeit requiring careful adaptation. Indeed, the emergence of mixed matter–photon states can challenge the theoretical foundations of these methodologies; for example, special transformations are needed to properly define molecular orbitals in QED environments \cite{Riso_2022}. Therefore, theories and models traditionally regarded as “axiomatic” in quantum chemistry and condensed matter physics cannot be assumed \textit{a priori}  in this context. Instead, they must be systematically extended and revalidated within the strong coupling regime. \par
\begin{figure}
    \centering
    \includegraphics[width=\linewidth]{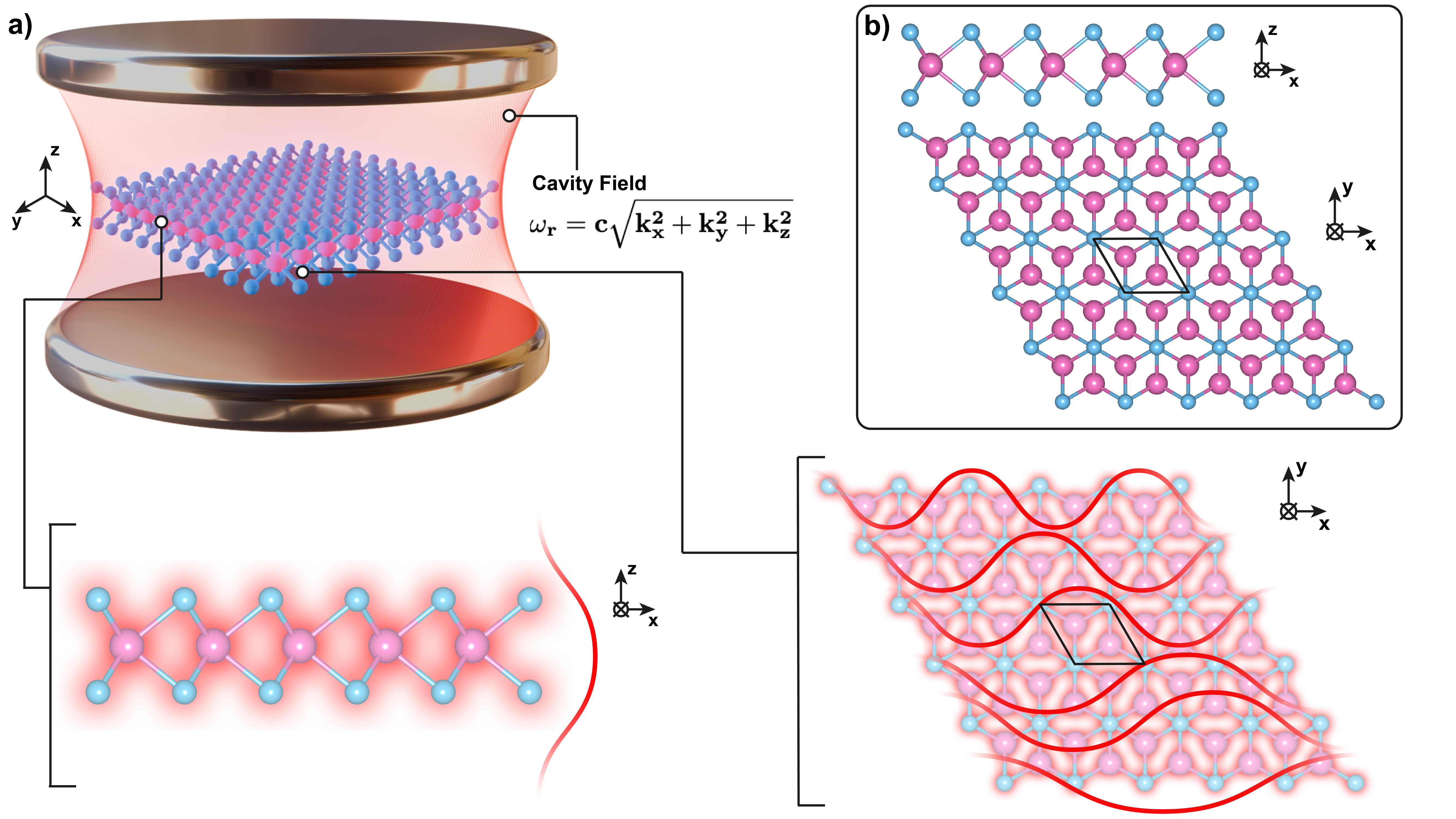}
    \caption{\textbf{Representation of a periodic material in a Fabry-Pérot cavity.} a) Two-dimensional periodic material in a cavity with resonance frequency $\omega_{\mathbf{r}}=c\sqrt{\mathbf{k_x}^2+\mathbf{k_y}^2+\mathbf{k_z}^2}$.
Along the cavity axis ($z$, lower left inset), the field is confined by the cavity, which justifies the single-mode approximation. Besides, its wavelength $\lambda_{z}=\frac{\pi}{k_z}$ is much longer than the material thickness, allowing the use of the long-wavelength approximation. In the cavity plane ($x$, $y$, lower right inset), the cavity is open, so that multiple in-plane photon modes must be considered and the long-wavelength approximation does not hold. 
b) Cross-sectional views along the cavity axis (top) and in the cavity plane (bottom) with the hexagonal unit cell indicated.}
    \label{fig:Fig1}
\end{figure}

At the core of this challenge lies a fundamental question, highlighted by Schlawin et al. \cite{Schlawin_2022} as one of the central open problems in the context of crystals in cavities: does Bloch's theorem—the cornerstone of condensed matter physics—remain valid when a crystal interacts with a quantized field inside a cavity? By defining the quantum states of electrons in a periodic lattice, Bloch's theorem underpins band theory and links microscopic periodicity to macroscopic properties. Yet, spatially structured cavity fields might in principle disrupt translational symmetry, leaving the theorem’s applicability uncertain. To date, to circumvent this issue, most of theoretical studies of crystals in cavities have employed the so-called long-wavelength approximation (LWA) which assumes the field as spatially uniform over atomic scale, thereby preserving the system's periodicity \cite{Svendsen_2025}. Although this simplification has provided valuable insights, including effective single-mode schemes that encode multimode effects and the restoration of translational symmetry in external magnetic fields \cite{Svendsen_2025,Rokaj_2019}, it is not formally justified for extended crystals, whose size can be comparable to the spatial variation of the field (Figure \ref{fig:Fig1}). A significant step beyond this paradigm was taken by Taylor et al. \cite{Taylor_2024,Taylor_2026}, who introduced a transformation of the QED Hamiltonian that renders the field spatially independent, thereby restoring Bloch’s theorem and leading to a formulation mathematically similar to the LWA.

Here, we take a different approach and show that Bloch’s theorem extends intrinsically to crystalline solids strongly coupled to spatially varying quantized fields beyond the LWA. The full light–matter Hamiltonian retains a generalized translational symmetry consistent with the lattice, enabling a polaritonic Bloch's theorem and the explicit construction of polaritonic Bloch functions.
 The key insight is that the quantized field acts locally on each electron, effectively as a one-body potential, thereby preserving the lattice periodicity. Moreover, we introduce a general framework to incorporate multimode cavity effects, weighting the field modes according to the Planck statistics. This analysis reveals that contributions beyond the characteristic cavity mode are finite at low frequencies and high temperatures.
Together, these advances provide a robust and physically consistent foundation for predictive studies of cavity-modified material properties and collective light–matter phenomena.

\subsection*{Extention of Bloch's theorem to cavity QED}

Bloch's theorem \cite{Bloch_1928} relies on the fundamental condition that the Hamiltonian is translationally invariant $T(\mathbf{R}) \hat{H} T^{\dagger}(\mathbf{R}) = \hat{H}$ for every Bravais lattice vector $\mathbf{R}$. This ensures that monoelectronic eigenstates can be expressed in the Bloch form as a plane wave times a function periodic in the lattice $\psi_{n\mathbf{q}(\mathbf{r})}=e^{i\mathbf{q}\cdot\mathbf{r}}u_{n\mathbf{q}}(\mathbf{r})$. \par 
However, in cavity QED where electrons are strongly coupled to quantized electromagnetic modes, the situation fundamentally changes. In fact, the non-relativistic Pauli–Fierz Hamiltonian in the Coulomb Gauge: 

\begin{equation} \hat{H}_{PF}=\frac{\Big(\hat{p}(\mathbf{r})-\hat{A}(\mathbf{r})\Big)^{2}}{2} + \sum_{I}V(\mathbf{r}-\mathbf{R}_{I}) + \sum_{\mathbf{k},\lambda}\omega_{k}(\hat{n}_{\mathbf{k},\lambda}+\frac{1}{2}) \label{eq:RES_BT1} \end{equation} 

\noindent does not satisfy the condition $T(\mathbf{R}) \hat{H}_{PF} T^{\dagger}(\mathbf{R}) = \hat{H}_{PF}$ under the ordinary electronic translation operator $\hat{T}_{e}(\mathbf{R})=e^{i\hat{p}\cdot\mathbf{R}}$. This breakdown originates from the vector potential $\hat{A}(\mathbf{r})= \sum_{\mathbf{k},\lambda}\sqrt{\frac{2\pi}{V\omega_{\mathbf{k}}}}\Bigl(\boldsymbol{\epsilon}_{\mathbf{k},\lambda} \hat{b}_{\mathbf{k},\lambda}e^{i\mathbf{k}\cdot\mathbf{r}} +\boldsymbol{\epsilon}^{*}_{\mathbf{k},\lambda}\hat{b}^{\dagger}_{\mathbf{k},\lambda}e^{-i\mathbf{k}\cdot\mathbf{r}} \Bigl)$ which acquires a phase $e^{\pm i\mathbf{k}\cdot\mathbf{R}}$ upon translation (see equation \ref{eq:SI_BT_4} of Supplementary Text S1). Consequently, the conjugated momentum $\hat{\pi}(\mathbf{r})=\hat{p}(\mathbf{r})-\hat{A}(\mathbf{r})$ and hence, the light-matter kinetic term $\frac{\Big(\hat{p}(\mathbf{r})-\hat{A}(\mathbf{r})\Big)^{2}}{2}$ ceases to be invariant under purely electronic translations. \par 
Nevertheless, translational invariance can be restored by introducing a global translation operator that acts jointly on the electronic and photonic degrees of freedom:

\begin{equation} \hat{T}_{glob}(\mathbf{R})=\hat{T}_{e}(\mathbf{R})\hat{T}_{ph}(\mathbf{R})=e^{i\hat{p}\cdot \mathbf{R}}e^{i\sum_{\mathbf{k},\lambda}\mathbf{k}\cdot\mathbf{R}\hat{n}_{\mathbf{k},\lambda}} \label{eq:RS_BT_4} 
\end{equation}

\noindent where $\sum_{\mathbf{k},\lambda}\mathbf{k}\hat{n}_{\mathbf{k},\lambda}$ is associated the total photonic momentum operator.
Indeed, under this combined operation, the Pauli–Fierz Hamiltonian recovers full translational invariance, $\hat{T}_{glob}(\mathbf{R})\hat{H}_{PF}\hat{T}_{glob}(\mathbf{R})^{\dagger}=\hat{H}_{PF}$. The existence of this symmetry is not merely formal: it reflects the fact that, in the strong light–matter coupling regime, electrons and photons no longer behave as distinct entities but as a single, hybrid quasiparticle—the polariton. \par 
Thus, electronic and photonic coordinates translate coherently, giving rise to polaritonic Bloch states whose collective symmetry underlies the extension of Bloch's theorem to cavity QED. The corresponding eigenstates can be written as: 
\begin{equation} \Psi_{n\mathbf{q}}(\mathbf{r},\{n_{\mathbf{k},\lambda}\}) = e^{i\mathbf{q}\cdot\mathbf{r}}\,u_{n\mathbf{q}}(\mathbf{r}) \sum_{\{n_{\mathbf{k},\lambda}\}} \prod_{\mathbf{k},\lambda} C^{\mathbf{k}}_{n_{\mathbf{k},\lambda}} \,e^{-i\mathbf{k}\cdot\mathbf{r}\,n_{\mathbf{k},\lambda}} (\hat{b}^{\dagger}_{\mathbf{k},\lambda})^{n_{\mathbf{k},\lambda}}|0\rangle \label{eq:RS_BT_3} \end{equation} 

\noindent where the photonic phases $e^{-i\mathbf{k}\cdot \mathbf{r}n_{\mathbf{k},\lambda}}$ ensure that the electronic quasimomentum $\mathbf{q}$ remains a good quantum number for all multiphotonon and multimode configurations. In this representation, the phase acquired by the electronic translation is exactly compensated by the photonic contribution, preserving periodicity through the global light–matter symmetry (See Supplementary Text S1).

\subsection*{Effect of the multimode field on the energetics of extended
materials}
The extension of Bloch's theorem to QED environments demonstrates that crystalline periodicity can persist under strong coupling, even beyond the LWA. This raises a natural question: how does the multimode structure of the cavity field affect the energetics of a crystal? In fact, in systems whose size is comparable to the spatial variation of the field, the continuum of the modes parallel to the mirrors surface, inherent to the open cavity, cannot be ignored and may introduce finite corrections to the energy.

To illustrate this effect, we consider a model system: a two-dimensional crystal in a Fabry–Pérot cavity with mirrors placed at $z=0$ and $L_{z}=0$. This geometry confines photons along the cavity axis ($z$-axis) while remaining open in the plane of the material (plane $xy$) (Figure \ref{fig:Fig1}).

We then decompose the vector potential into two components:

\begin{equation}
\hat{\mathbf{A}}(\mathbf{r}) = 
\underbrace{\hat{\mathbf{A}}_{k_z}(z,t)}_{k_x=k_y=0} \hspace{4pt}+ 
\underbrace{\hat{\mathbf{A}}_{\mathrm{obl}}(\mathbf{r},t)}_{k_z \text{ fixed},\, k_x,k_y \neq 0}.
\label{eq:RS_MM_1}
\end{equation}

\noindent The first is a resonant term along the cavity axis capturing the dominant light-matter interaction. 
The second is an oblique component representing the infinite set of oblique modes $(\pm k_x, \pm k_y, \pm k_z)$ that share the $z$-component of the wave vector with the resonant mode $\pm k_z$ but differ in their in-plane components $(k_x, k_y)$ (see Supplementary Text S2.1. and Figure \ref{fig:Fig2}).

Because the cavity is open in the in-plane directions $x$ and $y$, the oblique modes are treated in the limit of infinite in-plane extension ($L_{x} \xrightarrow{}+\infty$; $L_{y} \xrightarrow{}+\infty$), effectively forming a continuous spectrum. 

The photonic wave function is chosen within the single-photon approximation, a standard approach in cavity QED, where only vacuum and single-photon occupations of each mode are considered. We separate the wave function into a component $\Psi_{k_z}$ depending solely on the characteristic cavity mode $\pm k_z$ and an oblique component $\Psi_{\mathrm{obl}}$, corresponding to all in-plane wavevectors $\mathbf{k}_{\parallel}$ at fixed $|k_{z}|$. The full photonic state reads:

\begin{equation}
\begin{split}
\ket{\Psi_{\mathrm{Phot}}} &= 
\underbrace{
\Bigl(c_{|k_z|} (\hat{\beta}^\dagger_{\mathbf{0},1} + \hat{\beta}^\dagger_{\mathbf{0},2}) + \sqrt{1-2c^2_{|k_z|}}\Bigr)
}_{\Psi_{k_z}} \cdot \\&
\underbrace{
\Bigl(c_0 + \sum_{\mathbf{k}_{j} = (\mathbf{k}_{\parallel},|\mathbf{k}_z|)} c_{\mathbf{k}_j} e^{-i \mathbf{k}_{\parallel} \cdot \mathbf{r}_{\parallel} } (\hat{\alpha}^\dagger_{\mathbf{k}_j,1} + \hat{\beta}^\dagger_{\mathbf{k}_j,2})\Bigr)
}_{\Psi_{\mathrm{obl}}}
\ket{0_{\mathbf{k}_j}, 0_{\mathbf{k}_{j+1}}, \dots}
\label{eq:RS_MM_1}
\end{split}
\end{equation}

\noindent where $e^{-i\mathbf{k}_{\parallel}\cdot \mathbf{r}_{\parallel}}$ is the photonic phase of the polaritonic Bloch functions as shown in Equation \ref{eq:RS_BT_3}.
The analysis proceeds in two steps, applied to both the interaction ($\hat{\mathbf{p}}\!\cdot\!\hat{\mathbf{A}}$) and diamagnetic ($\hat{\mathbf{A}}^2$) terms of the Hamiltonian. 
In the first step, we evaluate the expectation value of the vector potential ($\hat{p}\cdot \langle \Psi_{\mathrm{obl}} | \hat{\mathbf{A}} | \Psi_{\mathrm{obl}} \rangle$) and of its square ($\langle \Psi_{\mathrm{obl}} | \hat{\mathbf{A}}^{2} | \Psi_{\mathrm{obl}} \rangle$) over the normalized oblique photon component. 
For both the expectation values only the oblique mode contributions survive (see Supplementary Text S2.4.).

In the second step, the resulting averaged field is combined with the resonant component $\hat{\mathbf{A}}_{k_z}$, and the expectation value is evaluated on $\Psi_{k_z}$, 
with the coefficient $c_{|k_z|}$ optimized to ensure normalization while incorporating the oblique-mode correction. 
Within this framework, the system can be interpreted as a single characteristic cavity mode $|k_z|$ dressed by a continuum of oblique modes, 
thereby isolating and quantifying the effect of the extended multimode field on the energetic landscape of the crystal.

Although this second step completes the full framework, in the present work we focus exclusively on the first stage.

The explicit form of $\Psi_{\mathrm{obl}}$ is crucial for the calculation of the expectation values. In this representation, the phase factor $e^{-i \mathbf{k}_{\parallel} \cdot \mathbf{r}_{\parallel}}$ originates from the polaritonic Bloch function in the plane of the crystal and depends only on the in-plane wavevector $\mathbf{k}_{\parallel} = (k_x, k_y)$, reflecting the two-dimensional periodicity of the material (see Equation \ref{eq:RS_BT_3}). The photonic ladder operators are defined as:
\begin{equation}
\hat{\alpha}_{\mathbf{k},\lambda=1,2} = \frac{\hat{b}_{\mathbf{k}_{\parallel}, +k_z, \lambda} + \hat{b}_{\mathbf{k}_{\parallel}, -k_z, \lambda}}{\sqrt{2}}, 
\qquad
\hat{\beta}_{\mathbf{k},\lambda=1,2} = \frac{\hat{b}_{\mathbf{k}_{\parallel}, +k_z, \lambda} - \hat{b}_{\mathbf{k}_{\parallel}, -k_z, \lambda}}{\sqrt{2}},
\label{eq:RS_MM_2}
\end{equation}

\noindent which correspond to standing–wave combinations of the $\pm k_z$ propagating cavity modes and are therefore independent of the sign of $k_z$.

The choice of $\hat{\alpha}$ and $\hat{\beta}$ operators and their polarizations is dictated by the cavity boundary conditions \cite{Svendsen_2025} which require the parallel electric field and the normal magnetic field to vanish at the mirrors. $\hat{\beta}_{\mathbf{0},\lambda=1/2}$ refer to resonant modes with $k_x=k_y=0$, while $\hat{\alpha}_{\mathbf{k}{j},\lambda=1/2}$ and $\hat{\beta}_{\mathbf{k}_{j},\lambda=1/2}$ represent the oblique ones with $(k_x,k_y)\neq 0$ at fixed $\pm k_z$ (see Supplementary Information S2.1.).

The coefficients $c_0$ and $c_{\mathbf{k}_j}$ set the statistical weight of each oblique mode in the photonic wave function. While many choices of amplitudes are in principle admissible—as long as normalization is preserved—we assign them according to thermal Planck statistics. This choice reflects the fact that the oblique field comprises an effectively infinite set of modes, for which a statistical description is the most physically reasonable. In fact, in the absence of external driving or other selection rules, temperature remains the only relevant scale governing their population. Accordingly, the oblique states are weighted by their Planck probabilities, such that the vacuum and single–photon contributions read:
\begin{subequations}
\begin{align}
&c_0 = \sqrt{P_{\mathrm{vac}}} = \sqrt{\prod_{\mathbf{k}} \left( 1 - e^{- \omega_{\mathbf{k}} /T} \right)} \\
&c_{\mathbf{k}_j} = \sqrt{P_{\mathrm{vac}}}\ e^{- \omega_{\mathbf{k}_j} / 2  T}
\end{align}
\end{subequations}

\noindent where $P_{\mathrm{vac}}$ denotes the probability that all oblique modes are unoccupied, and the product runs over all oblique wavevectors. However, once the single-photon approximation is relaxed, the multimode contribution acquires a  spatial dependence, so that the long-wavelength picture no longer applies, and additional multimode terms will naturally arise.

Interestingly, within the single photon approximation $\bra{\Psi_{\mathrm{obl}}} \hat{\mathbf{A}}_{\mathrm{obl}} \ket{\Psi_{\mathrm{obl}}}$ results spatially uniform in the crystal plane because the phase of the photonic Bloch function exactly cancels the in-plane wave dependence of the vector potential $\hat{\mathbf{A}}_{\mathrm{obl}}$ (see Materials and Methods M1 \cite{methods}).
Thus, the oblique field mirrors a long-wavelength-like behavior even when the coupling involves a multimode cavity field and a periodic system. This sheds light on why the long-wavelength approximation often remains qualitatively accurate in spatially extended architectures. 
Accordingly, within this approximation, our approach is qualitatively equivalent to that of Taylor et al. \cite{Taylor_2024,Taylor_2026}, yet retains the capability to capture spatially resolved multiphotonic field effects if more photonic states are considered.

Besides, as a result of this phase cancellation only the $z$-component of the coupling term $\hat{\mathbf{p}} \cdot \hat{\mathbf{A}}$ survives. 
This indicates that, although the field contains both in-plane and out-of-plane components, the effective coupling to the electronic momentum is mediated only through its $z$-component. In other words, the oblique modes contribute dominantly through their in-plane wavevector components $(k_x,k_y)$, which collectively generate an effective field polarized along $z$. This eveals a subtle decoupling between the characteristic cavity mode $k_z$ and the oblique modal continuum. Conversely, recovering the full multimode coupling—including in-plane components—requires a multiphoton description, where interference between distinct modes along $x$ and $y$ axis is preserved.   

Importantly, when the field is evaluated at the cavity center ($z = L_z/2$), corresponding to the long-wavelength limit along $z$, the contribution of the oblique multimode field vanishes completely if the single-photon approximation is also imposed because the latter contribution is proportional to $\cos(k_{z}z)$. Therefore, in this combined limit the multimode character of the field is effectively hidden from the observable light-matter interaction.

After establishing the theoretical picture, we quantify how the multimode photonic field modifies the energy landscape depending on temperature and cavity geometry. To this end, we compute the expectation values $\langle \Psi_{\mathrm{obl}} | \hat{p} \cdot \hat{\mathbf{A}}_{\mathrm{obl}} | \Psi_{\mathrm{obl}} \rangle$ and $\langle \Psi_{\mathrm{obl}} | \hat{\mathbf{A}}_{\mathrm{obl}}^2 | \Psi_{\mathrm{obl}} \rangle$ which result in integrals over the continuous manifold of in-plane modes $\mathbf{k}_{\parallel}$ (see Equations \ref{eq:MET_3} and \ref{eq:MET_4} in Materials and Methods M1 \cite{methods}).

In the limit of infinite in-plane dimensions, the normalization factor of these expectation values $\Tilde{A}_{0} \sim \sqrt{\frac{1}{2\pi L{z}}}$ depends only on the cavity length $L_{z}$. In contrast, the normalization of $\hat{\mathbf{A}}_{k_z}$ scales with the full cavity volume $V = L_x L_y L_z$. This difference reflects that, per unit length along $z$, the oblique modes contribute with a relatively larger prefactor compared to the single resonant mode. However, the actual contribution of the oblique modes to the total field ultimately depends on the integrals over the in-plane modes which can make it significantly smaller than that of the resonant mode. Moreover, these integrals are sensitive to temperature and cavity features (i.e. geometry and mirrors composition).

The integral associated with the coupling term $\hat{p} \cdot \hat{A}_{\mathrm{obl}}$ cannot be evaluated analytically but it is readily computed numerically (see Materials and Methods M2 \cite{methods}).
 Figure \ref{fig:Fig3} shows the behavior of the bare integral as a function of the cavity resonance frequency and the temperature. 
 
Panel a) shows the integral as a function of $k_z$ for both cryogenic ($T=100$ K) and ambient ($T=300$ K) conditions.
In both cases, the multimode contribution remains finite at low frequencies, with a clear enhancement at higher temperature consistent with the thermal population of oblique modes.
This contribution, however, is confined to the microwave and low-energy phonon region: as frequency increases toward the optical-phonon and electronic-excitation range, the integral rapidly vanishes.
Thus, multimode corrections are primarily relevant in the infrared-to-terahertz regime, where cavity fields can efficiently interact with collective low-energy excitations in solids.

Panel b) displays the same integral as a function of temperature for selected cavity resonance frequencies. 
In all cases, the integral grows monotonically with $T$, approaching saturation at high temperatures where the thermal occupation of oblique modes becomes substantial. 
This confirms that the strength of the multimode correction is thermally activated and that its magnitude can be tuned through both cavity design and environmental temperature.

Strikingly, in the low-frequency and high-temperature regime, the multimode integral reaches values comparable to the expectation of the resonant cavity mode. For instance, at a resonance of 10 GHz ($k_z \sim 10^{-8}$ a.u.), it ranges between $10^{-8}$ and $10^{-7}$ a.u. for temperatures from 100 K to 300 K. This means that it is of the same order as $\langle A_{k_z}\rangle$ when fully expressed, including the $1/\sqrt{k_z}$ factor and assuming $L_x = L_y = L_z$.

While the resonant contribution scales linearly with the cavity wavevector $A_{k_z} \propto k_z$, the oblique-mode correction combines the $\sqrt{k_z}$ prefactor with the multimode integral $\langle A_{\rm obl}\rangle \sim \sqrt{k_z}\, I$. Thus, in this regime where $I$ grows roughly linearly with $k_z$, one finds $\langle A_{\rm obl}\rangle \propto k_z^{3/2}$, showing that the multimode contribution is inherently significantly smaller than the resonant reference.

These observations indicate that, at the level of expectation values, the single-mode, long-wavelength term dominates, in line with common approximations used in molecular polaritonics, even for crystals in cavities. Remarkably, this dominance—and the apparent validity of the LWA in extended crystalline systems—emerges naturally within the single-photon approximation, as a direct consequence of the Bloch-like structure of the wave functions.

A similar trend was observed by Ying and co-workers \cite{Ying_2024}, who reported only minor contributions from off-resonant modes to vibrationally resolved rate constants, highlighting that multimode corrections can often be subleading. At the same time, a small expectation value of $\langle{\hat{A}_{\rm obl}}\rangle$ does not preclude a potentially significant impact on the material density of states once the coupled light-matter Schrödinger equation is fully solved, as shown by Ribeiro \cite{Ribeiro_2022}, who found substantial modifications in the density of states when including multiple modes. In this sense, our findings are complementary rather than contradictory: they demonstrate that a formally robust, parameter-free treatment can quantify the direct photonic contribution of oblique modes, while leaving open the possibility of non-negligible material response in a full multimode treatment. Crucially, this framework is fully general and readily compatible with periodic ab initio cavity QED methods, such as Hartree-Fock, density-functional and correlated approaches, enabling systematic exploration of multimode effects in realistic materials.

The expectation value of the diamagnetic term $\braket{\hat{\mathbf{A}}^{2}_{\mathrm{obl}}}$ is dominated by two divergent terms (see Equations \ref{eq:MET_5} and \ref{eq:MET_6} in Materials and Methods M1 \cite{methods}). The latter occur in the limit $k_{\parallel} \rightarrow +\infty$ and are independent of the cavity resonance frequency or the operating temperature. In other words, this is a genuine ultraviolet-type divergence arising solely from the multimode character of the cavity field, rather than from experimental conditions.

However, both integrals diverge as $\sqrt{k_{\parallel}^{2}+k_{z}^{2}}$. Consequently, even without applying the LWA along the cavity axis, the spatial dependence of the diamagnetic term vanishes. This is because the same divergent factor multiplies $\sin^{2}(k_{z}z)$ and $\cos^{2}(k_{z}z)$ , and their sum eliminates any 
$z$-dependence according to the fundamental trigonometric identity.

To regularize these integrals, we impose a cutoff at 
$k_{\parallel}$ such that $\frac{\sqrt{k^{2}_{\parallel}+k^{2}_{z}}}{c}=\omega_{p}$, where $\omega_{p}$
is the plasma frequency of the cavity mirrors. Indeed, frequencies above $\omega_{p}$ are no longer reflected by the mirrors and are dominated by the material dispersion; thus, they do not contribute physically to the cavity field. Typical plasma frequencies of metals used as mirrors are of the order of thousands THz—far above the cavity resonance frequencies, even when probing electronic excitations. This large separation of scales provides a clear physical basis for introducing this upper-frequency cutoff.

\subsection*{Discussion and Outlook}
In this work, we derive a polaritonic extension of Bloch’s theorem, revealing an intrinsic symmetry of crystalline systems under strong coupling. This symmetry is a combined light–matter translational invariance of the full polaritonic Hamiltonian, from which polaritonic Bloch functions follow directly. Within this framework, the crystal quasi-momentum is identified as the conserved quantity associated with combined translations, while lattice periodicity is preserved by polaritonic quasiparticles as the fundamental carriers of the crystal’s translational order.

We have also established a general framework to account for multimode cavity fields in crystalline systems. Within this approach, the dominant longitudinal cavity mode naturally emerges as the principal channel of light–matter interaction, while the continuum of oblique modes contributes only subleading corrections. In the single-photon regime, these contributions reduce to an effective field that is spatially uniform in the crystal plane, reinforcing the validity of the long-wavelength picture. Together, these results formally justify the single-mode and long-wavelength approximations widely employed in molecular polaritonics and demonstrate that they remain accurate and controlled in the solid-state limit.

Finally, this work establishes a rigorous theoretical foundation for predictive studies of cavity-modified materials, offering a pathway to explore temperature, geometry and mode-dependent effects in quantum materials, opening the door to controlled engineering of polaritonic phases in crystals.


\begin{figure}
    \centering
    \includegraphics[width=\linewidth]{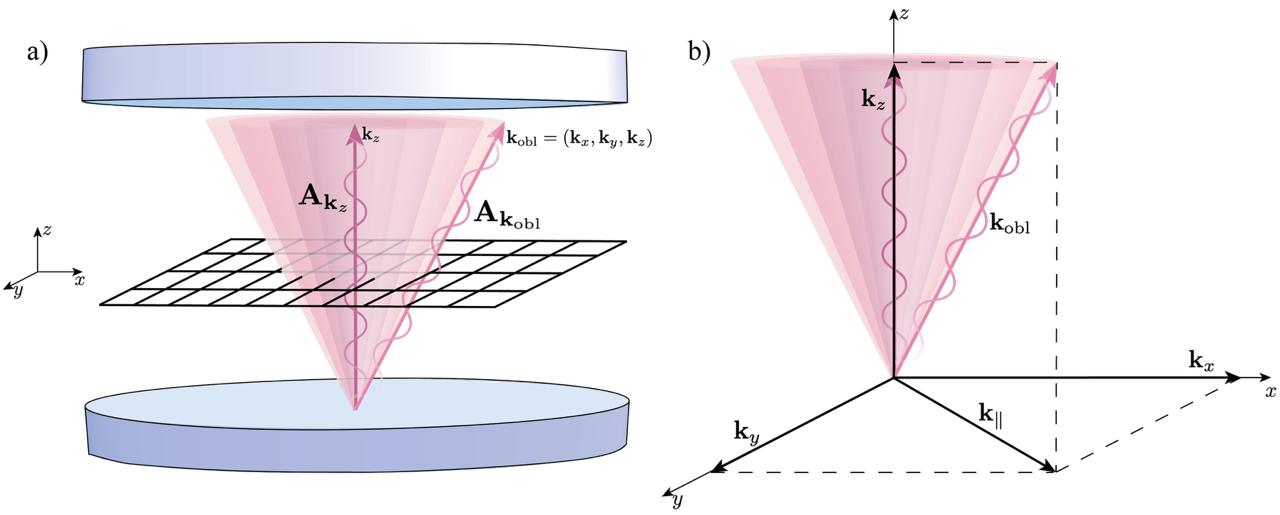}
    \caption{ \textbf{ Schematic representation of the cavity field.} a) The principal resonant component of the vector potential ($\hat{A}_{\mathbf{k}_z}$) lies at the center of a cone made of all the other possible oblique components $\hat{A}_{\mathbf{k}_{\mathrm{obl}}}$. b) Scheme of the in-plane ($\mathbf{k}_{\parallel}$) and out-of-plane ($\mathbf{k}_{z}$) components of the oblique modes of the cavity field. }
    \label{fig:Fig2}
\end{figure}

\begin{figure}
    \centering
    \includegraphics[width=\linewidth]{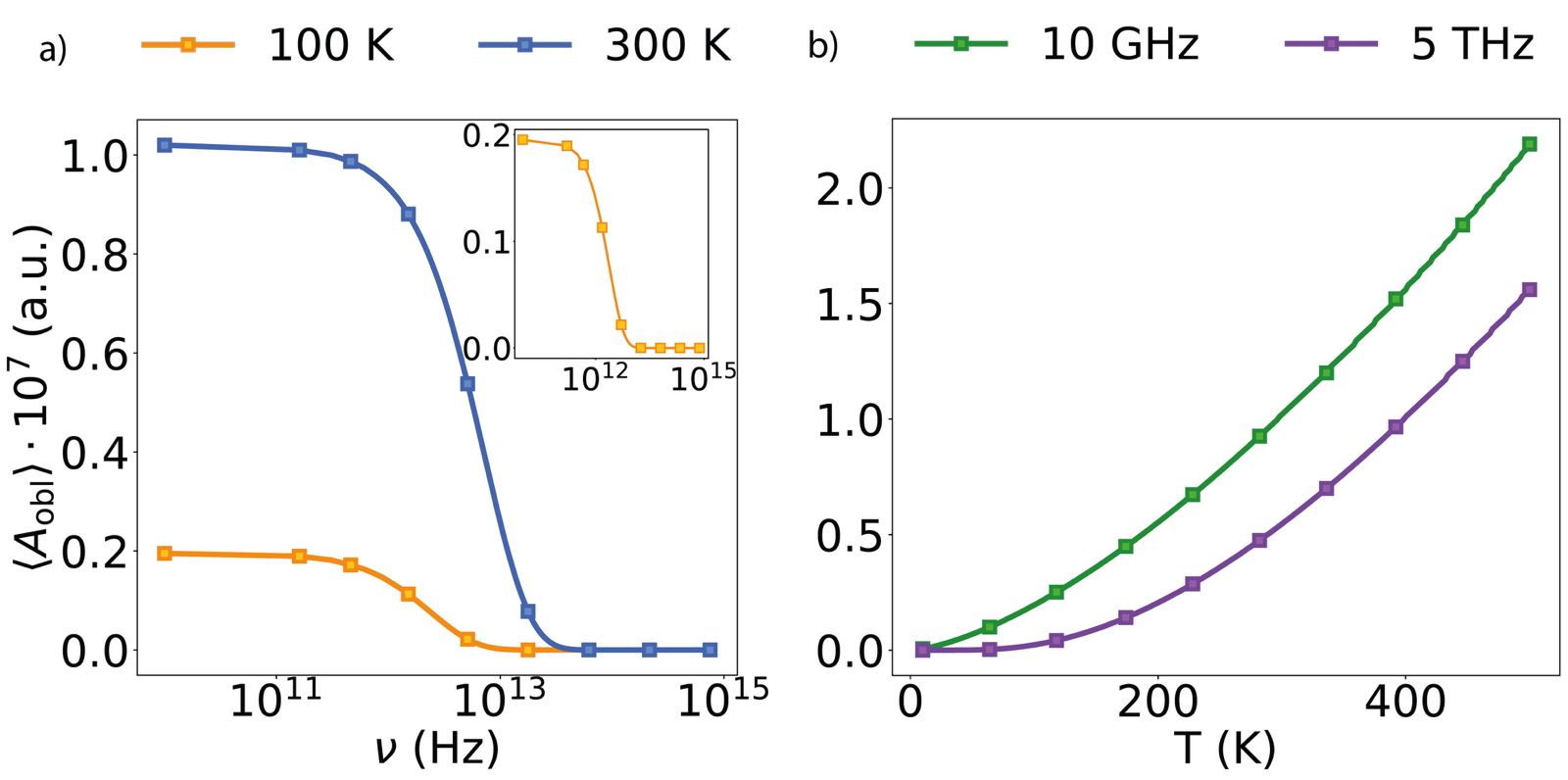}
    \caption{\textbf{Dependence of the multimode field integral on cavity and experimental parameters.} a) Bare integral (prior to multiplication by the prefactor $2\Tilde{A}_{0}\cos(k_{z}z)\,p_{z}\,P_{vac}$, see Equation \ref{eq:MET_3} in Methods \ref{subsec1_met}) versus cavity resonance frequency. b) Bare integral versus operating temperature. }
    \label{fig:Fig3}
\end{figure}


\clearpage 

%



\section*{Acknowledgments}
The authors thank M. Ruggenthaler, M. Kamper Svendsen, M.A.D. Taylor and P. Huo for fruitful discussions. G.B. and E.R. acknowledge funding from the
European Research Council (ERC) under the European
Union’s Horizon Europe Research and Innovation Programme (Grant n. ERC-StG-2021-101040197 - QED-SPIN). H.K. and R.R.R. acknowledge funding from the European Research Council (ERC) under the European Union’s Horizon 2020 Research and Innovation Programme (grant agreement No. 101020016).
\paragraph*{Funding:}
G.B. and E.R. were funded by the
European Research Council (ERC) under the European
Union’s Horizon Europe Research and Innovation Programme (Grant n. ERC-StG-2021-101040197 - QED-SPIN). H.K. and R.R.R. were funded by the European Research Council (ERC) under the European Union’s Horizon 2020 Research and Innovation Programme (grant agreement No. 101020016).
\paragraph*{Author contributions:}
E.R. conceived the project. G.B. and R.R.R. carried out the theoretical derivations. G.B. performed the numerical simulations. All authors participated in the discussion and
interpretations of the results. E.R. and H.K. supervised the project.
G.B. wrote the first draft and all authors discussed it. 
\paragraph*{Competing interests:}
There are no competing interests to declare.
\paragraph*{Data and materials availability:}
The Python codes used to evaluate the numerical integrals discussed in this article, together with usage instructions and example input and output files, is publicly available at https://doi.org/10.5281/zenodo.17880430 \cite{bruno_zenodo}.


\subsection*{Supplementary materials}
Materials and Methods\\
Supplementary Text\\


\newpage


\renewcommand{\thefigure}{S\arabic{figure}}
\renewcommand{\thetable}{S\arabic{table}}
\renewcommand{\theequation}{S\arabic{equation}}
\renewcommand{\thepage}{S\arabic{page}}
\setcounter{figure}{0}
\setcounter{table}{0}
\setcounter{equation}{0}
\setcounter{page}{1} 


\begin{center}
\section*{Supplementary Materials for\\ \scititle}

Giovanna Bruno$^{1\ast}$,
	Rosario R. Riso$^{2}$,
	Henrik Koch$^{2}$ and Enrico Ronca$^{1\ast}$
    \\
	\small$^{1}$Dipartimento di Chimica, Biologia e Biotecnologie, Universit\`a degli Studi di Perugia, Perugia \& 06123, Italy.\\
	\small$^{2}$Department of Chemistry, Norwegian University of Science and Technology, Trondheim \& 7491, Norway.\\
	\small$^\ast$Corresponding author. Email: giovanna.bruno@unipg.it;
           enrico.ronca@unipg.it.

\end{center}

\subsubsection*{This PDF file includes:}
Materials and Methods\\
Supplementary Text\\

\newpage


\subsection*{Materials and Methods}
\subsubsection*{M1. Formulation of the oblique modes integrals}\label{subsec1_met}

We consider the oblique vector potential $\hat{\mathbf{A}}_{\mathrm{obl}}(\mathbf{r},t)$ 
satisfying the cavity boundary conditions 
$\hat{\mathbf{n}} \wedge \hat{\mathbf{E}}(z=0/L_{z},x,y,t)=0$ and 
$\hat{\mathbf{n}} \cdot \hat{\mathbf{B}}(z=0/L_{z},x,y,t)=0$. Its explicit form is (see Supplementary Text S2):
\begin{equation}
\begin{split}
\hat{\mathbf{A}}_{\mathrm{obl}}(\mathbf{r},t)_{k_z \text{ fixed}, k_x,k_y \neq 0} &= 
A_{0}\sqrt{\frac{2}{c}}\Bigg\{
\sum_{\substack{\mathbf{k_{||}}=-\infty \\ \mathbf{k_{||}}\neq \mathbf{0}}}^{+\infty}
\frac{|\mathbf{k_{||}}|^{-1}}{(|\mathbf{k}_{||}|^{2} + k_z^2)^{3/4}}
\Biggl[
e^{i\mathbf{k_{||}}\cdot \mathbf{r_{||}}}
\begin{pmatrix}
-i k_x k_z \sin(k_z z) \\
-i k_y k_z \sin(k_z z) \\
|\mathbf{k}_{||}|^{2} \cos(k_z z)
\end{pmatrix}
\hat{\alpha}_{\mathbf{k},1}  \\
&\quad +
e^{-i\mathbf{k_{||}}\cdot \mathbf{r_{||}}}
\begin{pmatrix}
i k_x k_z \sin(k_z z) \\
i k_y k_z \sin(k_z z) \\
|\mathbf{k}_{||}|^{2} \cos(k_z z)
\end{pmatrix}
\hat{\alpha}^{\dagger}_{\mathbf{k},1}
\Biggr] \\
&\quad +
\frac{i|\mathbf{k_{||}}|^{-1}\sin(k_{z}z)}{(|\mathbf{k_{||}}|^{2}+ k_z^2)^{1/4}}
\begin{pmatrix}
-k_{y} \\ k_{x} \\ 0
\end{pmatrix}
\left(
e^{i\mathbf{k_{||}}\cdot \mathbf{r_{||}}}\hat{\beta}_{\mathbf{k},2} -
e^{-i\mathbf{k_{||}}\cdot \mathbf{r_{||}}}\hat{\beta}^{\dagger}_{\mathbf{k},2}
\right)
\Bigg\}.
\end{split}
\label{eq:MET_1}
\end{equation}

\noindent The corresponding oblique–mode wave function is constructed as:
\begin{equation}
\ket{\Psi_{\mathrm{obl}}}=\sqrt{P_{vac}}
\Biggl(
1+\sum_{\mathbf{k}_{j}=(\mathbf{k}_{\parallel},|\mathbf{k}_{z}|)}
e^{-i\mathbf{k}_{\parallel}\cdot\mathbf{r}_{\parallel}}
e^{-\frac{\hbar\omega_{\mathbf{k}_{j}}}{2k_{B}T}}
\left(\hat{\alpha}^{\dagger}_{\mathbf{k}_{j},1}+\hat{\beta}^{\dagger}_{\mathbf{k}_{j},2}\right)
\Biggr)
\ket{0_{\mathbf{k}_{j}},0_{\mathbf{k}_{j+1}},...,0}.
\label{eq:MET_2}
\end{equation}

\noindent Expectation values of the linear and quadratic field operators are then computed on this state. 
Each photon in the wave function carries an in-plane phase factor 
$e^{-i\mathbf{k}_{\parallel}\cdot \mathbf{r}_{\parallel}}$, while the vector potential carries 
$e^{\pm i\mathbf{k}_{\parallel}\cdot \mathbf{r}_{\parallel}}$ associated with the field operators. 
In the single-photon approximation, only terms in which a photon created by the wave function 
is annihilated by the field, or vice versa, contribute to 
$\langle \Psi_{\mathrm{obl}} | \hat{p}\cdot \hat{\mathbf{A}}_{\mathrm{obl}} | \Psi_{\mathrm{obl}} \rangle$. 
These contributions take the form
$\langle 0 | \hat{\alpha}_{\mathbf{k},1} | 1_{\mathbf{k},1} \rangle$ and $
\langle 1_{\mathbf{k},1} | \hat{\alpha}^\dagger_{\mathbf{k},1} | 0 \rangle,
$ and analogously for $\hat{\beta}_{\mathbf{k},2}$. Each matrix element involves exactly two phase factors, whose product is
$e^{\pm i\mathbf{k}_{\parallel}\cdot \mathbf{r}_{\parallel}} 
\, e^{\mp i\mathbf{k}_{\parallel}\cdot \mathbf{r}_{\parallel}} = 1$.  Consequently, the expectation value of vector potential results independent of the in-plane position.

For the diamagnetic term $\langle  \Psi_{\mathrm{obl}} | \hat{\mathbf{A}}_{\mathrm{obl}}^{2} | \Psi_{\mathrm{obl}} \rangle$, the surviving contributions 
involve two field operators, namely 
$
\hat{\alpha}^\dagger_{\mathbf{k},1} \hat{\alpha}_{\mathbf{k},1}$ and $
\hat{\alpha}_{\mathbf{k},1} \hat{\alpha}^\dagger_{\mathbf{k},1}$
(and the corresponding $\hat{\beta}$ terms). In this case, the phase factors cancel among themselves 
prior to contraction with the wave function, ensuring that this term
is also independent of $\mathbf{r}_{\parallel}$, although the mechanism differs from the previous case.

These expectation values can be expressed as integrals over the in-plane momentum. The linear term reads:
\begin{equation}
\langle \Psi_{\mathrm{obl}} | \hat{p} \cdot \hat{\mathbf{A}}_{\mathrm{obl}} | \Psi_{\mathrm{obl}} \rangle
= 2\Tilde{A}_{0}\cos(k_{z}z)\,p_{z}\,P_{vac}
\int_{0}^{+\infty} 2\pi\, dk_{\parallel}\,
\frac{k^{2}_{\parallel}\,e^{-\frac{ c}{2T}\sqrt{k^{2}_{\parallel}+k^{2}_{z}}}}
{(k^{2}_{\parallel}+k^{2}_{z})^{3/4}} ,
\label{eq:MET_3}
\end{equation}
\noindent while the quadratic term is:
\begin{equation}
\begin{split}
\langle \Psi_{\mathrm{obl}} | \hat{\mathbf{A}}_{\mathrm{obl}}^{2} | \Psi_{\mathrm{obl}} \rangle 
=&\; 8\pi\Tilde{A}^{2}_{0}P_{vac}\int_{0}^{+\infty}dk_{\parallel}
\frac{k^{2}_{z}\sin^{2}(k_{z}z)\,k_{\parallel}+\cos^{2}(k_{z}z)\,k^{3}_{\parallel}}
{(k^{2}_{\parallel}+k^2_{z})^{3/2}}
e^{-\frac{c}{T}\sqrt{k^{2}_{\parallel}+k^2_{z}}} \\
&+ 8\pi\Tilde{A}^{2}_{0}P_{vac}\int_{0}^{+\infty}dk_{\parallel}
\frac{\sin^{2}(k_{z}z)\,k_{\parallel}}
{\sqrt{k^{2}_{\parallel}+k^2_{z}}}
e^{-\frac{c}{T}\sqrt{k^{2}_{\parallel}+k^2_{z}}} \\
&+ 2\pi\Tilde{A}^{2}_{0}P_{vac}\int_{0}^{+\infty}dk_{\parallel}
\frac{k^{2}_{z}\sin^{2}(k_{z}z)\,k_{\parallel}+\cos^{2}(k_{z}z)\,k^{3}_{\parallel}}
{(k^{2}_{\parallel}+k^2_{z})^{3/2}} \\
&+ 2\pi\Tilde{A}^{2}_{0}P_{vac}\int_{0}^{+\infty}dk_{\parallel}
\frac{\sin^{2}(k_{z}z)\,k_{\parallel}}
{\sqrt{k^{2}_{\parallel}+k^2_{z}}}.
\end{split}
\label{eq:MET_4}
\end{equation}

\noindent The integrals associated with $\langle \Psi_{\mathrm{obl}} | \hat{\mathbf{A}}_{\mathrm{obl}}^2 | \Psi_{\mathrm{obl}} \rangle$ can be analitically solved. However, the terms:

\begin{subequations}  
\begin{equation}\label{eq:MET_5}
2\pi\Tilde{A}^{2}_{0}P_{vac}\cos^2(k_{z}z)\int_{0}^{+\infty}dk_{\parallel}\frac{
    k^{3}_{\parallel}}{\sqrt{(k^{2}_{\parallel}+k^2_{z})^3}}
    \end{equation} 
    
\begin{equation}\label{eq:MET_6}
2\pi\Tilde{A}^{2}_{0}P_{vac}\sin^2(k_{z}z)\int_{0}^{+\infty}dk_{\parallel}\frac{k_{\parallel}
    }{\sqrt{k^{2}_{\parallel}+k^2_{z}}}
\end{equation} 
\end{subequations}

\noindent do not converge, thus requiring to be regularized (see main text and Supplementary Text S2.5.).  

On the contrary, the integral associated with $\langle \Psi_{\mathrm{obl}} | \hat{p}\!\cdot\!\hat{\mathbf{A}}_{\mathrm{obl}} | \Psi_{\mathrm{obl}} \rangle$ does not admit any analytical solution, so that it can only be numerically evaluated. 

\subsubsection*{M2. Numerical evaluation of $\langle \Psi_{\mathrm{obl}} | \hat{p}\!\cdot\!\hat{\mathbf{A}}_{\mathrm{obl}} | \Psi_{\mathrm{obl}}\rangle$ integral} \label{subsec2_met}

We compute the radial integral of Equation \ref{eq:MET_3} numerically for each value of $k_{z}$ and $T$.

The integrand is sharply peaked around a finite in-plane momentum 
$k_{\parallel} = k_{\mathrm{max}}(k_{z},T)$ and decays exponentially at large 
$k_{\parallel}$. To ensure robust and unbiased convergence, we employed an 
adaptive two–step numerical strategy:

\begin{enumerate}
\item \textbf{Peak localization}  
A logarithmic grid in $k_{\parallel}\in[10^{-12},10^{-2}]$ is sampled to identify 
the region of maximal weight, followed by a local linear refinement to obtain 
an accurate estimate of the peak position $k_{\mathrm{max}}$.

\item \textbf{Adaptive symmetric integration}  
Starting from $k_{\mathrm{max}}$, the integral is accumulated symmetrically in 
$k_{\parallel}$ using \texttt{quad} from \textsc{SciPy}, advancing in steps of
$\Delta k = 10^{-7}$.  
The process terminates once the contribution of the latest increment drops below 
$10^{-15}$.
\end{enumerate}

The resulting integral is multiplied by $2\pi$ to recover the angular contribution 
of polar coordinates. Numerical uncertainty, obtained by summing the local quadrature errors returned by \texttt{quad}, remains below $10^{-5} \%$ of the integral value for all parameters considered. All computations were performed in 
Python~3 using \textsc{NumPy} and \textsc{SciPy}; the full analysis scripts are available in the online repository https://doi.org/10.5281/zenodo.17880430\cite{bruno_zenodo}.

\subsection*{Supplementary Text}
\subsubsection*{S1. Proof of the Bloch's Theorem in QED environments}
\label{sec:BT_dem}

The necessary condition of the Bloch's theorem is that the one-electron Hamiltonian is translationally invariant under all translations of the Bravais lattice, i.e.

\begin{equation}
\hat{T}_{el}(\mathbf{R})\hat{H}\hat{T}^{\dagger}_{el}(\mathbf{R})=\hat{H} \quad \forall\; \mathbf{R} \in \mathcal{L}= \left\{ \mathbf{R} \ \bigg| \ \mathbf{R} = n_1 \mathbf{a}_1 + n_2 \mathbf{a}_2 + n_3 \mathbf{a}_3, \ n_1,n_2,n_3 \in \mathbb{Z} \right\}
\label{eq:SI_BT_1}
\end{equation}

\noindent where $\hat{T}_{el}(\mathbf{R}) = e^{i \hat{\mathbf{p}} \cdot \mathbf{R}}$ is the electronic translation operator.
However, the Hamiltonian is not translationally invariant in the case of an electron coupled to a quantized electromagnetic field, where the system is described by the Pauli-Fierz Hamiltonian

\begin{equation}
\hat{H}_{\rm PF} = \frac{\big(\hat{\mathbf{p}}(\mathbf{r}) - \hat{\mathbf{A}}(\mathbf{r},t)\big)^{2}}{2} 
+ \hat{V}(\mathbf{R}) + \hat{V}(\mathbf{r}-\mathbf{R}) 
+ \sum_{\lambda=1}^{2} \sum_{\mathbf{k}} \omega_{\mathbf{k}} \Big( \hat{b}^{\dagger}_{\mathbf{k},\lambda} \hat{b}_{\mathbf{k},\lambda} + \frac{1}{2} \Big)
\label{eq:SI_BT_2}
\end{equation}

\noindent Here, $\hat{\mathbf{A}}(\mathbf{r},t)$ is the quantized vector potential representing the electromagnetic field

\begin{equation}
\hat{\mathbf{A}}(\mathbf{r},t) = \sum_{\lambda=1}^{2} \sum_{\mathbf{k}} \sqrt{\frac{2\pi}{V \omega_{\mathbf{k}}}} 
\Big( \boldsymbol{\epsilon}_{\mathbf{k},\lambda} \hat{b}_{\mathbf{k},\lambda} e^{i \mathbf{k} \cdot \mathbf{r}} 
+ \boldsymbol{\epsilon}^{*}_{\mathbf{k},\lambda} \hat{b}^{\dagger}_{\mathbf{k},\lambda} e^{-i \mathbf{k} \cdot \mathbf{r}} \Big)
\label{eq:SI_BT_3}
\end{equation}

\noindent and $\hat{b}^{\dagger}_{\mathbf{k},\lambda}$, $\hat{b}_{\mathbf{k},\lambda}$ are the creation and annihilation operators for the photonic mode $(\mathbf{k},\lambda)$. 

\noindent We note that the vector potential $\hat{\mathbf{A}}(\mathbf{r},t)$ is not translationally invariant under electronic translations. Indeed, applying the translation operator we obtain

\begin{align}
\hat{\mathbf{A}}(\mathbf{r} + \mathbf{R},t) &=\hat{T}_{el}(\mathbf{R}) \, \hat{\mathbf{A}}(\mathbf{r},t) \, \hat{T}^{\dagger}_{el}(\mathbf{R}) 
\\
&=  \sum_{\lambda=1}^{2} \sum_{\mathbf{k}} \sqrt{\frac{2\pi}{V \omega_{\mathbf{k}}}} 
\Big( \boldsymbol{\epsilon}_{\mathbf{k},\lambda} \hat{b}_{\mathbf{k},\lambda} e^{i \mathbf{k} \cdot \mathbf{r}}e^{i \mathbf{k} \cdot \mathbf{R}}
+ \boldsymbol{\epsilon}^{*}_{\mathbf{k},\lambda} \hat{b}^{\dagger}_{\mathbf{k},\lambda} e^{-i \mathbf{k} \cdot \mathbf{r}}e^{-i \mathbf{k} \cdot \mathbf{R}} \Big) \nonumber\neq \hat{\mathbf{A}}(\mathbf{r},t)
\label{eq:SI_BT_4}
\end{align}

\noindent and consequently the Pauli-Fierz Hamiltonian $\hat{H}_{\rm PF}$ does not satisfy eq. \ref{eq:SI_BT_1}. 

\noindent Nonetheless, the Pauli--Fierz Hamiltonian satisfies a more general translational invariance as it remains invariant under the combined translation of both electronic and photonic degrees of freedom, expressed via the global translation operator 

\begin{equation}
\hat{T}_{\mathrm{glob}}(\mathbf{R}) = \hat{T}_{\mathrm{el}}(\mathbf{R}) \, \hat{T}_{\mathrm{ph}}(\mathbf{R}) = e^{i\hat{\mathbf{p}}\cdot \mathbf{R}}e^{i \sum_{\mathbf{k}',\lambda} \mathbf{k}' \cdot \mathbf{R} \, \hat{n}_{\mathbf{k}',\lambda} }
\label{eq:SI_BT_5}
\end{equation}

\noindent where $\hat{n}_{\mathbf{k}',\lambda} = \hat{b}^{\dagger}_{\mathbf{k}',\lambda} \hat{b}_{\mathbf{k}',\lambda}$. The electronic and photonic translation operators each modify the phase factor on the photonic creation and annihilation operators, such that

\begin{equation}
\begin{split}
    \hat{T}_{glob}\hat{A}(\mathbf{r},t)\hat{T}^{\dagger}_{glob}&=\sqrt{\frac{2\pi}{V}}\sum_{\mathbf{k},\lambda}\sqrt{\frac{1}{\omega_{\mathbf{k}}}}\Big(\epsilon_{\mathbf{k},\lambda}e^{i\hat{\mathbf{p}}\cdot\mathbf{R}}e^{i\mathbf{k}\cdot\mathbf{r}}e^{-i\hat{\mathbf{p}}\cdot\mathbf{R}}e^{i \sum_{\mathbf{k}',\lambda} \mathbf{k}' \cdot \mathbf{R} \, \hat{n}_{\mathbf{k}',\lambda}}\hat{b}_{\mathbf{k},\lambda}e^{-i \sum_{\mathbf{k}',\lambda} \mathbf{k}' \cdot \mathbf{R} \, \hat{n}_{\mathbf{k}',\lambda}} + \\& \epsilon^{*}_{\mathbf{k},\lambda}e^{i\hat{\mathbf{p}}\cdot\mathbf{R}}e^{-i\mathbf{k}\cdot\mathbf{r}}e^{-i\hat{\mathbf{p}}\cdot\mathbf{R}}e^{i \sum_{\mathbf{k}',\lambda} \mathbf{k}' \cdot \mathbf{R} \, \hat{n}_{\mathbf{k}',\lambda}}\hat{b}^{\dagger}_{\mathbf{k},\lambda}e^{-i \sum_{\mathbf{k}',\lambda} \mathbf{k}' \cdot \mathbf{R} \, \hat{n}_{\mathbf{k}',\lambda}}\Big)
\end{split}
\label{eq:SI_BT_6}
\end{equation}

\noindent Employing the Baker-Campbell-Hausdorff (BCH) formula, we find

\begin{equation}
\begin{split}
e^{i\hat{\mathbf{p}}\cdot\mathbf{R}}e^{i\mathbf{k}\cdot\mathbf{r}}e^{-i\hat{\mathbf{p}}\cdot\mathbf{R}} =& e^{i\mathbf{k}\cdot\mathbf{r}}e^{i\mathbf{k}\cdot\mathbf{R}}\\
e^{i\hat{\mathbf{p}}\cdot\mathbf{R}}e^{-i\mathbf{k}\cdot\mathbf{r}}e^{-i\hat{\mathbf{p}}\cdot\mathbf{R}} =& e^{-i\mathbf{k}\cdot\mathbf{r}}e^{-i\mathbf{k}\cdot\mathbf{R}}
\label{eq:SI_BT_7}
\end{split}
\end{equation}
and
\begin{equation}
\begin{split}
e^{i \sum_{\mathbf{k}',\lambda} \mathbf{k}' \cdot \mathbf{R} \, \hat{n}_{\mathbf{k}',\lambda}}\hat{b}_{\mathbf{k},\lambda}e^{-i \sum_{\mathbf{k}',\lambda} \mathbf{k}' \cdot \mathbf{R} \, \hat{n}_{\mathbf{k}',\lambda}} =& e^{-i \mathbf{k} \cdot \mathbf{R}}\hat{b}_{\mathbf{k},\lambda} \\
e^{i \sum_{\mathbf{k}',\lambda} \mathbf{k}' \cdot \mathbf{R} \, \hat{n}_{\mathbf{k}',\lambda}}\hat{b}^{\dagger}_{\mathbf{k},\lambda}e^{-i \sum_{\mathbf{k}',\lambda} \mathbf{k}' \cdot \mathbf{R} \, \hat{n}_{\mathbf{k}',\lambda}} =& e^{i \mathbf{k} \cdot \mathbf{R}}\hat{b}^{\dagger}_{\mathbf{k},\lambda} 
\label{eq:SI_BT_8}
\end{split}
\end{equation}
such that the phase introduced by the electronic translation is exactly cancelled by the photonic contribution. As a result

\begin{equation}
    \hat{T}_{\mathrm{glob}}(\mathbf{R}) \, \hat{H}_{\mathrm{PF}} \, \hat{T}_{\mathrm{glob}}^\dagger(\mathbf{R}) = \hat{H}_{\mathrm{PF}}
\label{eq.SI_BT_9}
\end{equation}

\noindent This implies that the global translation operator $\hat{T}_{glob}(\mathbf{R})$ and the Pauli-Fierz Hamiltonian $\hat{H}_{PF}$ possess a common set of eigenfunctions.

\noindent To derive the common form of these functions, we start from the eigenfunctions of $\hat{T}_{el}(\mathbf{R})$ and $\hat{T}_{ph}(\mathbf{R})$ which fulfill the following eigenvalue equations

\begin{equation}
    \hat{T}_{el}(\mathbf{R})e^{i\mathbf{q}\cdot\mathbf{r}}=e^{i\mathbf{q}\cdot\mathbf{R}}e^{i\mathbf{q}\cdot\mathbf{r}}
\label{eq:SI_BT_10}
\end{equation}
\begin{equation}
    \hat{T}_{ph}(\mathbf{R})\prod_{\mathbf{k},\lambda}({\hat{b}}^{\dagger}_{\mathbf{k},\lambda})^{n_{\mathbf{k},\lambda}}\ket{0} = \prod_{\mathbf{k},\lambda}e^{in_{\mathbf{k},\lambda}\mathbf{k}\cdot\mathbf{R}}({\hat{b}}^{\dagger}_{\mathbf{k},\lambda})^{n_{\mathbf{k},\lambda}}\ket{0}=e^{i\sum_{\mathbf{k},\lambda}n_{\mathbf{k},\lambda}\mathbf{k}\cdot\mathbf{R}}\prod_{\mathbf{k},\lambda}({\hat{b}}^{\dagger}_{\mathbf{k},\lambda})^{n_{\mathbf{k},\lambda}}\ket{0}
\label{eq:SI_BT_11}
\end{equation}
\noindent where $e^{i\mathbf{q}\cdot \mathbf{R}}$ is the usual plane wave while $({\hat{b}}^{\dagger}_{\mathbf{k},\lambda})^{n_{\mathbf{k},\lambda}}\ket{0}$ is a multimode eigenfunction of the photonic translation operator for a fixed number of photons $n_{\mathbf{k},\lambda}$. \par

\noindent Therefore, a generic multimode and multiphoton eigenfunction of the global translation operator can be written as:
\begin{equation}
\begin{split}
\Psi_{el,ph}(\mathbf r)
&= \sum_{n_{\mathbf{k}_{1},1}}\sum_{n_{\mathbf{k}_{1},{2}}}\sum_{n_{\mathbf{k}_{2},{1}}}...\sum_{n_{\mathbf{k}_{N},{2}}}c_{n_{\mathbf{k}_{1},{1}}...n_{\mathbf{k}_{N},{2}}}e^{i\mathbf{q}\cdot \mathbf{r}}\prod^{\infty}_{j=1}\prod^{2}_{\lambda=1}
e^{-in_{\mathbf{k}_{j},\lambda}\mathbf{k}_{j} \cdot \mathbf r}
\big(\hat b^\dagger_{\mathbf{k}_{j},\lambda_{i}}\big)^{n_{\mathbf{k}_{j},\lambda}}
|0\rangle\\
&=\sum_{\{n_{\mathbf k,\lambda}\}}
c_{\{n_{\mathbf k,\lambda}\}}e^{i\mathbf{q} \cdot \mathbf r}
\prod_{\mathbf k,\lambda}
e^{-i n_{\mathbf{k},\lambda}\mathbf k \cdot \mathbf r}
\big(\hat b^\dagger_{\mathbf k,\lambda}\big)^{n_{\mathbf k,\lambda}}
|0\rangle
\label{eq:SI_BT_12}
\end{split}
\end{equation}

\noindent Applying $\hat T_{\mathrm{glob}}(\mathbf R)$ to this state gives
\begin{align}
\hat T_{\mathrm{glob}}(\mathbf R)\Psi_{el,ph}(\mathbf r)
&=\sum_{\{n_{\mathbf k,\lambda}\}}
c_{\{n_{\mathbf k,\lambda}\}}e^{i\mathbf{q}\cdot\mathbf{r}}e^{i\mathbf{q}\cdot \mathbf{R}}
\prod_{\mathbf k,\lambda}
e^{-in_{\mathbf k,\lambda}\mathbf k \cdot \mathbf r}
e^{-in_{\mathbf k,\lambda}\mathbf k \cdot \mathbf R}
e^{in_{\mathbf k,\lambda}\mathbf k \cdot \mathbf R}
\big(\hat b^\dagger_{\mathbf k,\lambda}\big)^{n_{\mathbf k,\lambda}}
|0\rangle \nonumber\\
&= e^{i\mathbf q\cdot\mathbf R}\,\Psi_{el,ph}(\mathbf r)
\label{eq:SI_BT_13}
\end{align}

\noindent Hence, $\Psi_{el,ph}(\mathbf{r})$ is an eigenfunction of the global translation operator with eigenvalue $e^{i\mathbf q\cdot\mathbf R}$.  
Note that the spatial phase factor $e^{-i n_{\mathbf k,\lambda}\mathbf k \cdot\mathbf r}$ ensures that the conserved crystal momentum of the coupled light-matter system coincides with the electronic crystal momentum $\mathbf q$ regardless the number of photons interacting with the electron.  
This property is essential because within the minimal-coupling framework the photonic state must allow superpositions of different photon numbers, such as the vacuum and one-photon states, while preserving the translational symmetry.

\noindent Thus, the polaritonic Bloch function reads

\begin{equation}
\begin{split}
    \Psi^{B}_{el,ph}(\mathbf r)
= &\Bigg(\sum_{\{n_{\mathbf k,\lambda}\}}
c_{\{n_{\mathbf k,\lambda}\}}e^{i\mathbf{q}\cdot \mathbf{r}}
\prod_{\mathbf k,\lambda}
e^{-in_{\mathbf k,\lambda}\mathbf{k}\cdot \mathbf r}
\big(\hat b^\dagger_{\mathbf k,\lambda}\big)^{n_{\mathbf k,\lambda}}\Bigg)\times\\
&\sum_{\mathbf{G}}c_{\mathbf{q}-\mathbf{G}}\Bigg(\sum_{\{m_{\mathbf k,\lambda}\}}
c_{\{m_{\mathbf k,\lambda}\}}
e^{-i\mathbf{G}\cdot \mathbf r}
\prod_{\mathbf{k},\lambda}e^{-im_{\mathbf k,\lambda}\mathbf{k} \cdot \mathbf r}\big(\hat b^\dagger_{\mathbf k,\lambda}\big)^{m_{\mathbf k,\lambda}}\Bigg)
|0\rangle
\label{eq:SI_BT_14}
\end{split}
\end{equation}
where $\mathbf{G}=n_{1}\mathbf{b}_{1}+n_{2}\mathbf{b}_{2}+n_{3}\mathbf{b}_{3}$ is a reciprocal lattice vector, while the function:

\begin{equation}
\sum_{\mathbf{G}}c_{\mathbf{q}-\mathbf{G}}\Bigg(\sum_{\{m_{\mathbf k,\lambda}\}}
c_{\{m_{\mathbf k,\lambda}\}}
e^{-i\mathbf{G}\cdot \mathbf r}
\prod_{\mathbf{k},\lambda}e^{-im_{\mathbf k,\lambda}\mathbf{k} \cdot \mathbf r}\big(\hat b^\dagger_{\mathbf k,\lambda}\big)^{m_{\mathbf k,\lambda}}\Bigg)
\label{eq:SI_BT_15}
\end{equation}
is an eigenfunction of $\hat{T}_{glob}(\mathbf{R})$ with eigenvalue 1, which means it corresponds to the periodic part of the polaritonic Bloch function. Moreover, by rearranging equation \ref{eq:SI_BT_14} we obtain

\begin{equation}
     \Psi^{B}_{el,ph}(\mathbf r) = e^{i\mathbf{q}\cdot\mathbf{r}}\sum_{\mathbf{G}}c_{\mathbf{q}-\mathbf{G}}e^{-i\mathbf{G}\cdot\mathbf{r}} \sum_{\{t_{\mathbf{k},\lambda}\}}\prod_{\mathbf{k},\lambda}e^{-it_{\mathbf{k},\lambda}\mathbf{k}\cdot\mathbf{r}}c_{t_{\mathbf{k}},\lambda}\big(\hat b^\dagger_{\mathbf k,\lambda}\big)^{t_{\mathbf k,\lambda}}
|0\rangle
\label{eq:SI_BT_16}
\end{equation}
Therefore, $\Psi^{B}_{el,ph}(\mathbf r)$ can be conveniently expressed as $\Psi^{B}_{el,ph}(\mathbf r) = \Psi_{el}(\mathbf r) \times \Psi_{ph}(\mathbf r)$, where $\Psi_{el}(\mathbf r)=e^{i\mathbf{q}\cdot\mathbf{r}}\sum_{\mathbf{G}}c_{\mathbf{q}-\mathbf{G}}e^{-i\mathbf{G}\cdot\mathbf{r}}$ is the usual electronic Bloch function, while $\Psi_{ph}(\mathbf r) =\sum_{\{t_{\mathbf{k},\lambda}\}}\prod_{\mathbf{k},\lambda}e^{-it_{\mathbf{k},\lambda}\mathbf{k}\cdot\mathbf{r}}c^{\mathbf{k}}_{t_{\mathbf{k}}}\big(\hat b^\dagger_{\mathbf k,\lambda}\big)^{t_{\mathbf k,\lambda}}\ket{0} $ is a general multimode and multiphoton wave function, with $t_{\mathbf{k},\lambda}=n_{\mathbf{k},\lambda}+m_{\mathbf{k},\lambda}$ being the number of photons and $e^{-it_{\mathbf{k},\lambda}\mathbf{k}\cdot\mathbf{r}}$ the phase allowing for the conservation of the total crystal momentum $\mathbf{q}$.

 \subsubsection*{S2. Quantum Statistical Modeling of the Cavity Field}
 \paragraph{S2.1. Disentanglement of Characteristic Cavity Modes from Oblique Modes in the Description of the Cavity Field}
 Under the Coulomb Gauge $\nabla \cdot \hat{\mathbf{A}}(\mathbf{r},t) = 0$, i.e. $\mathbf{k} \cdot \mathbf{\epsilon}_{\mathbf{k}} = 0$, and assuming the field to be linearly polarized, the free space vector potential is expressed as:
 
\begin{equation}
\hat{\mathbf{A}}(\mathbf{r},t)=A_{0}\sum_{\lambda=1}^{2}\sum_{\mathbf{k}=-\infty}^{+\infty}\frac{\boldsymbol{\epsilon}_{\mathbf{k},\lambda}}{\sqrt{\omega_{\mathbf{k}}}}\Bigl( \hat{b}_{\mathbf{k},\lambda}e^{i\mathbf{k}\cdot\mathbf{r}}  +\hat{b}^{\dagger}_{\mathbf{k},\lambda}e^{-i\mathbf{k}\cdot\mathbf{r}} \Bigl)  
\label{eq:SI_QS_1}
\end{equation}
where $A_{0}=\sqrt{\frac{2\pi}{V}}$ is the normalization factor.
Setting the cavity axis along $z$ and adopting the single-mode approximation in this direction, the characteristic cavity modes are $\pm\mathbf{k_z}=(0,0,\pm k_z)$. 
Accordingly, a generic mode of the cavity field can be expressed as:

\begin{equation}
  \mathbf{k} = \mathbf{k_{||}}\pm\mathbf{ k_{z}}
    \label{eq:SI_QS_2}
\end{equation}
where $\mathbf{k_{||}} = (k_x, k_y, 0)$ lies in the xy plane orthogonal to the cavity axis. These modes $\mathbf{k}_{\parallel} \pm \mathbf{k}_{z}$ are referred to as oblique modes, since the component $k_{z}$ is fixed by the cavity resonance $\omega_{\mathrm{res}} = c|k_{z}|$ and never vanishes, while only the in-plane component $\mathbf{k}_{\parallel}$ is free to vary.\par 
Under these assumptions, equation \ref{eq:SI_QS_1} can be reformulated as:

\begin{equation}
\begin{split}  
\hat{\mathbf{A}}(x,y,z,t) &= \sum_{\substack{\mathbf{k_{||}}=-\infty }}^{+\infty}  
 \frac{A_{0}}{c^{1/2} (|\mathbf{k_{||}}|^2  + k_z^2)^{1/4}} \Bigg\{e^{ik_{z}z}
 \biggl[
\Bigl( \hat{b}_{\mathbf{k_{||}}, k_z, 1}\epsilon_{\mathbf{k_{||}}, k_z, 1} + \hat{b}_{\mathbf{k_{||}}, k_z, 2}\epsilon_{\mathbf{k_{||}}, k_z, 2} \Bigr) 
e^{i\mathbf{k_{||}}\cdot\mathbf{r_{||}}}+ \\&\Bigl( \hat{b}^{\dagger}_{\mathbf{k_{||}}, -k_z, 1}\epsilon_{\mathbf{k_{||}}, -k_z, 1} + \hat{b}^{\dagger}_{\mathbf{k_{||}}, -k_z, 2}\epsilon_{\mathbf{k_{||}}, -k_z, 2}\Bigr) 
e^{-i\mathbf{k_{||}}\cdot\mathbf{r_{||}}} 
\biggr]+\\
&e^{-ik_{z}z} 
 \biggl[
\Bigl( \hat{b}_{\mathbf{k_{||}}, -k_z, 1}\epsilon_{\mathbf{k_{||}}, -k_z, 1} + \hat{b}_{\mathbf{k_{||}}, -k_z, 2}\epsilon_{\mathbf{k_{||}}, -k_z, 2} \Bigr) 
e^{i\mathbf{k_{||}}\cdot\mathbf{r_{||}}}+\\
&\Bigl( \hat{b}^{\dagger}_{\mathbf{k_{||}}, k_z, 1}\epsilon_{\mathbf{k_{||}}, k_z, 1} + \hat{b}^{\dagger}_{\mathbf{k_{||}}, k_z, 2}\epsilon_{\mathbf{k_{||}}, k_z, 2}\Bigr) 
e^{-i\mathbf{k_{||}}\cdot\mathbf{r_{||}}} 
\biggr]\Bigg\}
\label{eq:SI_QS_3}
\end{split}
\end{equation}
where $|\mathbf{k}_{||}|^{2}=k_{x}^2+k_{y}^2$  and $\mathbf{r_{||}} = (x, y,0)$ is the in-plane position vector. \par
Recasting the creation and annihilation operators in terms of more physical meaningful standing waves as:

\begin{equation}
\hat{\alpha}_{\mathbf{k},\lambda=1,2}=\frac{\hat{b}_{\mathbf{k_{||}},+k_{z},\lambda}+\hat{b}_{\mathbf{k_{||}},-k_{z},\lambda}}{\sqrt{2}} \hspace{1cm} \hat{\beta}_{\mathbf{k},\lambda=1,2}=\frac{\hat{b}_{\mathbf{k_{||}},+k_{z},\lambda}-\hat{b}_{\mathbf{k_{||}},-k_{z},\lambda}}{\sqrt{2}}
\label{eq:SI_QS_4}  
\end{equation}
\par
equation \ref{eq:SI_QS_3} can be further simplified:

\begin{equation}
\begin{aligned}
 \hat{\mathbf{A}}(x,y,z,t) &= 
 \sum_{\substack{\mathbf{k_{||}}=-\infty}}^{+\infty}  
\frac{A_{0}}{\sqrt{2c}(|\mathbf{k}_{||}|^{2} + k_z^2)^{1/4}} \Bigg\{e^{ik_{z}z}\Biggl[  
\Bigl( (\hat{\alpha}_{\mathbf{k},1} +\hat{\beta}_{\mathbf{k},1}) \epsilon_{\mathbf{k_{||}},k_z,1}+(\hat{\alpha}_{\mathbf{k},2} +\hat{\beta}_{\mathbf{k},2}) \epsilon_{\mathbf{k_{||}},k_z,2}\Bigr)e^{i\mathbf{k_{||}}\cdot\mathbf{r_{||}}} +\\
&\Bigl( (\hat{\alpha}^{\dagger}_{\mathbf{k},1} -\hat{\beta}^{\dagger}_{\mathbf{k},1}) \epsilon_{\mathbf{k_{||}},-k_z,1}+(\hat{\alpha}^{\dagger}_{\mathbf{k},2} -\hat{\beta}^{\dagger}_{\mathbf{k},2}) \epsilon_{\mathbf{k_{||}},-k_z,2}\Bigr)e^{-i\mathbf{k_{||}}\cdot\mathbf{r_{||}}}\Biggr]+\\&e^{-ik_{z}z}\Biggl[\Bigl( (\hat{\alpha}_{\mathbf{k},1} -\hat{\beta}_{\mathbf{k},1}) \epsilon_{\mathbf{k_{||}},-k_z,1}+(\hat{\alpha}_{\mathbf{k},2} -\hat{\beta}_{\mathbf{k},2}) \epsilon_{\mathbf{k_{||}},-k_z,2}\Bigr)e^{i\mathbf{k_{||}}\cdot\mathbf{r_{||}}}+\\&\Bigl( (\hat{\alpha}^{\dagger}_{\mathbf{k},1} +\hat{\beta}^{\dagger}_{\mathbf{k},1}) \epsilon_{\mathbf{k_{||}},k_z,1}+(\hat{\alpha}^{\dagger}_{\mathbf{k},2} +\hat{\beta}^{\dagger}_{\mathbf{k},2}) \epsilon_{\mathbf{k_{||}},k_z,2}\Bigr)e^{-i\mathbf{k_{||}}\cdot\mathbf{r_{||}}}\Biggr]\Bigg\}
\label{eq:SI_QS_5}
\end{aligned}
\end{equation}
As a preliminary step in developing the quantum statistical description of the cavity field, we separate the vector potential into two distinct contributions: the component associated with the fundamental longitudinal cavity mode, ${\hat{\mathbf{A}}_{k_z}(z,t)}_{k_x=k_y=0}$ and the complementary contribution $\hat{\mathbf{A}}_{\mathrm{obl}}(\mathbf{r},t)_{k_z \text{ fixed},\, k_x,k_y \neq 0}$, accounting for the infinite set of oblique modes with non-vanishing in-plane momentum (see Fig.1 in the main text). Accordingly:
\begin{equation}
    \hat{\mathbf{A}}(x,y,z,t)
    =
    {\hat{\mathbf{A}}_{k_z}(z,t)}_{k_x=k_y=0}
    +
    \hat{\mathbf{A}}_{\mathrm{obl}}(\mathbf{r},t)_{k_z \text{ fixed},\, k_x,k_y \neq 0}
    \label{eq:SI_QS_6}
\end{equation}
By separating the two contributions, equation \ref{eq:SI_QS_5} becomes:

\begin{equation}
\begin{split}
 \hat{\mathbf{A}}(x,y,z,t) &= 
\frac{A_{0}}{\sqrt{2c|k_{z}|}} \Bigg\{e^{ik_{z}z}\biggl[  
 (\hat{\alpha}_{\mathbf{0},1} +\hat{\beta}_{\mathbf{0},1}) \epsilon_{\mathbf{0},k_z,1}+(\hat{\alpha}_{\mathbf{0},2} +\hat{\beta}_{\mathbf{0},2}) \epsilon_{\mathbf{0},k_z,2}+\\
& (\hat{\alpha}^{\dagger}_{\mathbf{0},1} -\hat{\beta}^{\dagger}_{\mathbf{0},1}) \epsilon_{\mathbf{0},-k_z,1}+(\hat{\alpha}^{\dagger}_{\mathbf{0},2} -\hat{\beta}^{\dagger}_{\mathbf{0},2}) \epsilon_{\mathbf{0},-k_z,2}\biggr]+\\&e^{-ik_{z}z}\Biggl[(\hat{\alpha}_{\mathbf{0},1} -\hat{\beta}_{\mathbf{0},1}) \epsilon_{\mathbf{0},-k_z,1}+(\hat{\alpha}_{\mathbf{0},2} -\hat{\beta}_{\mathbf{0},2}) \epsilon_{\mathbf{0},-k_z,2}+\\&(\hat{\alpha}^{\dagger}_{\mathbf{0},1} +\hat{\beta}^{\dagger}_{\mathbf{0},1}) \epsilon_{\mathbf{0},k_z,1}+(\hat{\alpha}^{\dagger}_{\mathbf{0},2} +\hat{\beta}^{\dagger}_{\mathbf{0},2}) \epsilon_{\mathbf{0},k_z,2}\Biggr]\Bigg\}+\\&\sum_{\substack{\mathbf{k_{||}}=-\infty \\\mathbf{k_{||}}\ne \mathbf{0}}}^{+\infty}  
\frac{A_{0}}{\sqrt{2c}(|\mathbf{k}_{||}|^{2} + k_z^2)^{1/4}}\Bigg\{e^{ik_{z}z}\Biggl[\biggl((\hat{\alpha}_{\mathbf{k_{||}},1} +\hat{\beta}_{\mathbf{k_{||}},1})\epsilon_{\mathbf{k_{||}},k_z,1}+(\hat{\alpha}_{\mathbf{k_{||}},2} +\hat{\beta}_{\mathbf{k_{||}},2})\epsilon_{\mathbf{k_{||}},k_z,2}\biggr)e^{i\mathbf{k_{||}}\cdot\mathbf{r_{||}}}+\\&\biggl((\hat{\alpha}^{\dagger}_{\mathbf{k_{||}},1} -\hat{\beta}^{\dagger}_{\mathbf{k_{||}},1})\epsilon_{\mathbf{k_{||}},-k_z,1}+(\hat{\alpha}^{\dagger}_{\mathbf{k_{||}},2} -\hat{\beta}^{\dagger}_{\mathbf{k_{||}},2})\epsilon_{\mathbf{k_{||}},-k_z,2}\biggr)e^{-i\mathbf{k_{||}}\cdot\mathbf{r_{||}}}\Biggr]+\\&e^{-ik_{z}z}\Biggl[\biggl((\hat{\alpha}_{\mathbf{k_{||}},1} -\hat{\beta}_{\mathbf{k_{||}},1})\epsilon_{\mathbf{k_{||}},-k_z,1}+(\hat{\alpha}_{\mathbf{k_{||}},2} -\hat{\beta}_{\mathbf{k_{||}},2})\epsilon_{\mathbf{k_{||}},-k_z,2}\biggr)e^{i\mathbf{k_{||}}\cdot\mathbf{r_{||}}}+\\&\biggl((\hat{\alpha}^{\dagger}_{\mathbf{k_{||}},1} +\hat{\beta}^{\dagger}_{\mathbf{k_{||}},1})\epsilon_{\mathbf{k_{||}},k_z,1}+(\hat{\alpha}^{\dagger}_{\mathbf{k_{||}},2} +\hat{\beta}^{\dagger}_{\mathbf{k_{||}},2})\epsilon_{\mathbf{k_{||}},k_z,2}\biggr)e^{-i\mathbf{k_{||}}\cdot\mathbf{r_{||}}}\Biggr]\Bigg\}
\label{eq:SI_QS_7}
\end{split}
\end{equation}
where $\hat{\alpha}_{\mathbf{0},\lambda=1,2}(\hat{\beta}_{\mathbf{0},\lambda=1,2})$ and $\hat{\alpha}^{\dagger}_{\mathbf{0},\lambda=1,2}(\hat{\beta}^{\dagger}_{\mathbf{0},\lambda=1,2})$ are the bosonic operators associated with the cavity modes $(0,0,\pm k_{z})$ and $\epsilon_{\mathbf{0},\pm k_{z},\lambda=1,2}$ are the corresponding polarization versors. \par
By imposing:

\begin{equation}
\begin{split}
\boldsymbol{\epsilon}_{\mathbf{k_{||}},\pm k_{z},1}=\frac{1}{|\mathbf{k_{||}}|\sqrt{|\mathbf{k}_{||}|^{2}+k_{z}^{2}}}\begin{pmatrix}
\mp k_{x}k_{z} \\
\mp k_{y}k_{z}\\
|\mathbf{k}_{||}|^{2}
\end{pmatrix}; \hspace{1cm}\boldsymbol{\epsilon}_{\mathbf{k_{||}},\pm k_{z},2}&=\frac{1}{|\mathbf{k_{||}}|}\begin{pmatrix} -k_{y} \\  k_{x} \\0\end{pmatrix}\\ \\
\boldsymbol{\epsilon}_{\mathbf{0},\pm k_{z},1}=\begin{pmatrix}
   1 \\
   0 \\
   0
\end{pmatrix}; \hspace{1cm}
\boldsymbol{\epsilon}_{\mathbf{0},\pm k_{z},2}=\begin{pmatrix}
   0 \\
   1 \\
   0
\end{pmatrix}
\label{eq:SI_QS_8}
\end{split}
\end{equation}
where $|\mathbf{k_{||}}|=\sqrt{k_{x}^2+k_{y}^2}$, 
and exploiting the Euler identity $e^{\pm ik_{z}z}=\cos(k_{z}z)\pm  i \sin(k_{z}z)$, the global vector potential takes the form:

\begin{equation}
\begin{split}
\hat{\mathbf{A}}(x,y,z,t) &= A_{0}\sqrt{\frac{2}{c|k_z|}} \Biggl[
\begin{pmatrix}
    1 \\
    0 \\
    0
\end{pmatrix}
\left(
\cos(k_z z)(\hat{\alpha}_{\mathbf{0},1} + \hat{\alpha}^\dagger_{\mathbf{0},1}) 
+ i \sin(k_z z)(\hat{\beta}_{\mathbf{0},1} + \hat{\beta}^\dagger_{\mathbf{0},1})
\right) +\\& 
\begin{pmatrix}
    0 \\
    1 \\
    0
\end{pmatrix}
\left(
\cos(k_z z)(\hat{\alpha}_{\mathbf{0},2} + \hat{\alpha}^\dagger_{\mathbf{0},2}) 
+ i \sin(k_z z)(\hat{\beta}_{\mathbf{0},2} + \hat{\beta}^\dagger_{\mathbf{0},2})
\right) \Biggr] + \\
& \sum_{\substack{\mathbf{k_{||}}=-\infty \\ \mathbf{k_{||}}\neq \mathbf{0}}}^{+\infty}
A_0\sqrt{\frac{2}{c|\mathbf{k}_{||}|^{2}}} \frac{1}{(|\mathbf{k}_{||}|^{2} + k_z^2)^{3/4}} \Biggl[
e^{i\mathbf{k_{||}}\cdot \mathbf{r_{||}}}  \Biggl(
\begin{pmatrix}
    -i k_x k_z \sin(k_z z) \\
    -i k_y k_z \sin(k_z z) \\
    |\mathbf{k}_{||}|^{2} \cos(k_z z)
\end{pmatrix} \hat{\alpha}_{\mathbf{k},1} +\\
&\begin{pmatrix}
    -k_x k_z \cos(k_z z) \\
    -k_y k_z \cos(k_z z) \\
    i|\mathbf{k}_{||}|^{2} \sin(k_z z)
\end{pmatrix} \hat{\beta}_{\mathbf{k},1}
\Biggr)+e^{-i\mathbf{k_{||}}\cdot \mathbf{r_{||}}}\Biggl( \begin{pmatrix}
    i k_x k_z \sin(k_z z) \\
    i k_y k_z \sin(k_z z) \\
    |\mathbf{k}_{||}|^{2} \cos(k_z z)
\end{pmatrix}\hat{\alpha}^{\dagger}_{\mathbf{k},1}+\\
&\begin{pmatrix}
    -k_x k_z \cos(k_z z) \\
   - k_y k_z \cos(k_z z) \\
   -i |\mathbf{k}_{||}|^{2} \sin(k_z z)
\end{pmatrix}\hat{\beta}^{\dagger}_{\mathbf{k},1}\Biggr)\Biggr]+\\&\sum_{\substack{\mathbf{k_{||}}=-\infty \\ \mathbf{k_{||}} \neq \mathbf{0}}}^{+\infty}
A_0\sqrt{\frac{2 }{c|\mathbf{k}_{||}|^{2}}} \frac{1}{(|\mathbf{k}_{||}|^{2}+ k_z^2)^{1/4}} \Biggl[\begin{pmatrix}
    -k_{y} \\
    k_{x} \\
    0
\end{pmatrix} \Biggl(cos(k_{z}z)(e^{i\mathbf{k_{||}}\cdot \mathbf{r_{||}}}\hat{\alpha}_{\mathbf{k},2}+e^{-i\mathbf{k_{||}}\cdot \mathbf{r_{||}}}\hat{\alpha}^{\dagger}_{\mathbf{k},2})+\\
&i \sin(k_{z}z)(e^{i\mathbf{k_{||}}\cdot \mathbf{r_{||}}}\hat{\beta}_{\mathbf{k},2}-e^{-i\mathbf{k_{||}}\cdot \mathbf{r_{||}}}\hat{\beta}^{\dagger}_{\mathbf{k},2})\Biggr)\Biggr]
\label{eq:SI_QS_9}
\end{split}
\end{equation}

\paragraph{S2.2. Boundary Conditions for the Cavity Field}
In this work we enforce the standard boundary conditions at the cavity mirrors, requiring the electric field  ($\hat{\mathbf{E}}$) to be normal and the magnetic field ($\hat{\mathbf{B}}$) to be tangential to the cavity boundaries $z=0$ and $z=L_{z}$, namely:

\begin{subequations}\label{eq:SI_QS_10}
\begin{align}
\hat{\mathbf{n}} \wedge \hat{\mathbf{E}}(z=0/L_{z},x,y,t) &= 0 \label{eq:SI_QS_10a} \\
\hat{\mathbf{n}} \cdot \hat{\mathbf{B}}(z=0/L_{z},x,y,t) &= 0 \label{eq:SI_QS_10b}
\end{align}
\end{subequations}

where $\hat{\mathbf{n}}$ is the versor normal at the surfaces $z=0$ and $z=L_{z}$. By assuming the Coulomb Gauge to the free field, the electric and magnetic fields result mutually orthogonal:

\begin{equation}
\begin{split}
     &\hat{\mathbf{E}}= -\partial_t \hat{\mathbf{A}} \\
     & \hat{\mathbf{B}}=\nabla \times \hat{\mathbf{A}}
\end{split}
\label{eq:SI_QS_11}
\end{equation}
So that the constraint \ref{eq:SI_QS_10a} implies the constraint \ref{eq:SI_QS_10b} and viceversa. Thus, for the sake of conciseness, herein we derive the correct expression of the vector potential $\hat{\mathbf{A}}$ using only the first of the two constraints. \par
By imposing the quantization condition $k_{z} = \frac{n_{z}\pi}{L_{z}}$ with $n_{z} \in \mathbb{N}$ on the characteristic cavity modes $(0,0,\pm k_{z})$, the boundary conditions \ref{eq:SI_QS_10a} and \ref{eq:SI_QS_10b} are automatically satisfied at the mirror surface $z=L_{z}$.\par
As for the boundary at $z=0$, the corresponding electric field reads:

\begin{equation}
\begin{split}
\hat{\mathbf{E}}(x,y,z=0,t) &= iA_{0}\sqrt{2c|k_z|}\Biggl[\begin{pmatrix}
    1 \\
    0 \\
    0
\end{pmatrix}(\hat{\alpha}_{\mathbf{0},1}-\hat{\alpha}^{\dagger}_{\mathbf{0},1})+\begin{pmatrix}
    0 \\
    1 \\
    0
\end{pmatrix}(\hat{\alpha}_{\mathbf{0},2}-\hat{\alpha}^{\dagger}_{\mathbf{0},2})\Biggr]+\\
&\sum_{\substack{\mathbf{k_{||}}=-\infty \\ \mathbf{k_{||}}\neq \mathbf{0}}}^{+\infty}
i A_{0}\sqrt{\frac{2 c}{|\mathbf{k}_{||}|^{2}}} \frac{1}{(|\mathbf{k}_{||}|^{2} + k_z^2)^{1/4}} \Biggl[e^{i\mathbf{k_{||}}\cdot \mathbf{r_{||}}}\Biggl(\begin{pmatrix}
    0 \\
    0 \\
    |\mathbf{k}_{||}|^{2}
\end{pmatrix}\hat{\alpha}_{\mathbf{k},1} + \begin{pmatrix}
    -k_{x}k_{z} \\
    -k_{y}k_{z} \\
    0
\end{pmatrix}\hat{\beta}_{\mathbf{k},1}\Biggr) +\\
&e^{-i\mathbf{k_{||}}\cdot \mathbf{r_{||}}}\Biggl(\begin{pmatrix}
    0 \\
    0 \\
    -|\mathbf{k}_{||}|^{2}
\end{pmatrix}\hat{\alpha}^{\dagger}_{\mathbf{k},1}+\begin{pmatrix}
    k_{x}k_{z} \\
    k_{y}k_{z} \\
    0
\end{pmatrix}\hat{\beta}^{\dagger}_{\mathbf{k},1}\Biggr)\Biggr]+\\
&iA_{0}\sqrt{\frac{2c}{|\mathbf{k}_{||}|^{2}}}(|\mathbf{k}_{||}|^{2}+k_{z}^2)^{1/4}\Biggl[\begin{pmatrix}
    -k_{y} \\
    k_{x} \\
    0
\end{pmatrix}(e^{i\mathbf{k_{||}}\cdot \mathbf{r_{||}}}\hat{\alpha}_{\mathbf{k},2}-e^{-i\mathbf{k_{||}}\cdot \mathbf{r_{||}}}\hat{\alpha}^{\dagger}_{\mathbf{k},2})\Biggr]
\label{eq:SI_QS_12}
\end{split}
\end{equation}

\noindent To find the correct expression of the vector potential, we make the constraint \ref{eq:SI_QS_10a} more stringent by imposing:

\begin{equation}
    \bra{\Psi_{Phot}}|\hat{\mathbf{E}}(x,y,z=0,t)|^{2}\ket{\Psi_{Phot}}\equiv 0 
    \label{eq:SI_QS_13}
\end{equation}
with $\ket{\Psi_{Phot}}$ being a generic photonic wave function. Since the choice of the latter is entirely arbitrary, herein we employ the vacuum state 
$\ket{0}$ in order to simplify the derivation.
The only non null terms are associated with the expectation values on the vacuum state $\ket{0}$ of the number operators $\alpha_{\mathbf{0},\lambda=1,2}\alpha^{\dagger}_{\mathbf{0},\lambda'=1',2'}$, $\alpha_{\mathbf{k'},\lambda=1,2}\alpha^{\dagger}_{\mathbf{k},\lambda'=1',2'}$ and $\beta_{\mathbf{k'},\lambda=1,2}\beta^{\dagger}_{\mathbf{k},\lambda'=1',2'}$, which implies $\mathbf{k'} \equiv \mathbf{k}$, $1' \equiv 1$, $2' \equiv 2$ and $t' \equiv t$. The result is:

\begin{equation}
    \bra{0}|\hat{\mathbf{E}}(x,y,0,t)|^{2}\ket{0} = 2cA_{0}^{2}\biggl(2|k_{z}| + \sum_{\substack{\mathbf{k_{||}}=-\infty \\ \mathbf{k_{||}} \neq \mathbf{0}}}^{+\infty}\frac{k_{z}^2}{\sqrt{|\mathbf{k}_{||}|^{2}+k_{z}^2}}+\sqrt{|\mathbf{k}_{||}|^{2}+k_{z}^2}\biggr)
\label{eq:SI_QS_14}
\end{equation}

\noindent The latter is a positive definite function of the field modes; therefore, only the terms identically null at $z=0$ have to be included in the expression of the vector potential:

\begin{equation}
    \begin{split}
\hat{\mathbf{A}}(x,y,z,t) &= A_{0}\sqrt{\frac{2}{c}}\Bigg\{ \frac{i\sin(k_{z}z)}{\sqrt{|k_{z}|}}\Biggl[
\begin{pmatrix}
    1 \\
    0 \\
    0
\end{pmatrix}
(\hat{\beta}_{\mathbf{0},1} + \hat{\beta}^\dagger_{\mathbf{0},1})
 + 
\begin{pmatrix}
    0 \\
    1 \\
    0
\end{pmatrix}
(\hat{\beta}_{\mathbf{0},2} + \hat{\beta}^\dagger_{\mathbf{0},2}) \Biggr] + \\
&\sum_{\substack{\mathbf{k_{||}}=-\infty \\ \mathbf{k_{||}}\neq \mathbf{0}}}^{+\infty}
 \frac{|\mathbf{k_{||}}|^{-1}}{(|\mathbf{k}_{||}|^{2} + k_z^2)^{3/4}} \Biggl[
e^{i\mathbf{k_{||}}\cdot \mathbf{r_{||}}}  
\begin{pmatrix}
    -i k_x k_z \sin(k_z z) \\
    -i k_y k_z \sin(k_z z) \\
    |\mathbf{k}_{||}|^{2} \cos(k_z z)
\end{pmatrix} \hat{\alpha}_{\mathbf{k},1}+e^{-i\mathbf{k_{||}}\cdot \mathbf{r_{||}}} \begin{pmatrix}
    i k_x k_z \sin(k_z z) \\
    i k_y k_z \sin(k_z z) \\
    |\mathbf{k}_{||}|^{2} \cos(k_z z)
\end{pmatrix}\hat{\alpha}^{\dagger}_{\mathbf{k},1}\Biggr]+\\&\frac{i|\mathbf{k_{||}}|^{-1}\sin(k_{z}z)}{(|\mathbf{k}_{||}|^{2}+ k_z^2)^{1/4}} \begin{pmatrix}
    -k_{y} \\
    k_{x} \\
    0
\end{pmatrix} (e^{i\mathbf{k_{||}}\cdot \mathbf{r_{||}}}\hat{\beta}_{\mathbf{k},2}-e^{-i\mathbf{k_{||}}\cdot \mathbf{r_{||}}}\hat{\beta}^{\dagger}_{\mathbf{k},2})\Bigg\}        
    \end{split}
\label{eq:SI_QS_15}
\end{equation}

\noindent where
\begin{equation}
    \hat{\mathbf{A}}_{k_z}(z,t)_{k_x=k_y=0}= A_{0}\sqrt{\frac{2}{c}} \frac{i\sin(k_{z}z)}{\sqrt{|k_{z}|}}\Biggl[
\begin{pmatrix}
    1 \\
    0 \\
    0
\end{pmatrix}
(\hat{\beta}_{\mathbf{0},1} + \hat{\beta}^\dagger_{\mathbf{0},1})
 + 
\begin{pmatrix}
    0 \\
    1 \\
    0
\end{pmatrix}
(\hat{\beta}_{\mathbf{0},2} + \hat{\beta}^\dagger_{\mathbf{0},2}) \Biggr]
\label{eq:SI_QS_16}
\end{equation}
\noindent and
\begin{equation}
    \begin{split}
   \hat{\mathbf{A}}_{\mathrm{obl}}(\mathbf{r},t)_{k_z \text{ fixed},\, k_x,k_y \neq 0}&= A_{0}\sqrt{\frac{2}{c}}\Bigg\{\sum_{\substack{\mathbf{k_{||}}=-\infty \\ \mathbf{k_{||}}\neq \mathbf{0}}}^{+\infty}
 \frac{|\mathbf{k_{||}}|^{-1}}{(|\mathbf{k}_{||}|^{2} + k_z^2)^{3/4}} \cdot \\& \Biggl[
e^{i\mathbf{k_{||}}\cdot \mathbf{r_{||}}}  
\begin{pmatrix}
    -i k_x k_z \sin(k_z z) \\
    -i k_y k_z \sin(k_z z) \\
    |\mathbf{k}_{||}|^{2} \cos(k_z z)
\end{pmatrix} \hat{\alpha}_{\mathbf{k},1}+e^{-i\mathbf{k_{||}}\cdot \mathbf{r_{||}}} \begin{pmatrix}
    i k_x k_z \sin(k_z z) \\
    i k_y k_z \sin(k_z z) \\
    |\mathbf{k}_{||}|^{2} \cos(k_z z)
\end{pmatrix}\hat{\alpha}^{\dagger}_{\mathbf{k},1}\Biggr]+\\& \frac{i|\mathbf{k_{||}}|^{-1}\sin(k_{z}z)}{(|\mathbf{k}_{||}|^{2}+ k_z^2)^{1/4}} \begin{pmatrix}
    -k_{y} \\
    k_{x} \\
    0
\end{pmatrix} (e^{i\mathbf{k_{||}}\cdot \mathbf{r_{||}}}\hat{\beta}_{\mathbf{k},2}-e^{-i\mathbf{k_{||}}\cdot \mathbf{r_{||}}}\hat{\beta}^{\dagger}_{\mathbf{k},2})\Bigg\}
\end{split}
\label{eq:SI_QS_17}
\end{equation}

\paragraph{S2.3. Continuum Limit of the In-plane Cavity Modes}
Since the cavity is open along the directions \(x\) and \(y\), the corresponding field modes are treated in the limit \(L_x \to \infty\) and \(L_y \to \infty\). In this limit, the discrete sums over the transverse wavevectors,

\begin{equation}
    \sum_{\substack{\mathbf{k}_{\parallel}=-\infty \\ \mathbf{k}_{\parallel}\neq \mathbf{0}}}^{+\infty}
=
\sum_{\substack{k_{x}=-\infty \\ k_{x}\neq 0}}^{+\infty}
\sum_{\substack{k_{y}=-\infty \\ k_{y}\neq 0}}^{+\infty}
\label{eq:SI_QS_18}
\end{equation}

\noindent appearing in equation \ref{eq:SI_QS_17}, turn into integrals according to the continuum substitution:

\begin{equation}
 \sum_{{k}_{x}}\sum_{{k}_{y}} \;\longrightarrow\; 
\frac{L_x L_y}{(2\pi)^2}\int \mathrm{d}{k}_{x} \mathrm{d}{k}_{y}   
\label{eq:SI_QS_19}
\end{equation}

\noindent Thus, the oblique-mode contribution to the vector potential becomes:
\begin{equation}
\begin{split}
   \hat{\mathbf{A}}_{\mathrm{obl}}(\mathbf{r},t)_{k_z \text{ fixed},\, k_x,k_y \neq 0}
   &= \tilde{A}_{0}\sqrt{\frac{2}{c}}
   \Bigg\{
   \int_{-\infty}^{+\infty}\mathrm{d}k_x
   \int_{-\infty}^{+\infty}\mathrm{d}k_y\,
   \frac{|\mathbf{k}_{\parallel}|^{-1}}{(|\mathbf{k}_{\parallel}|^{2} + k_z^2)^{3/4}} \cdot \\
   &\Biggl[
   e^{i\mathbf{k}_{\parallel}\cdot\mathbf{r}_{\parallel}}
   \begin{pmatrix}
    -i k_x k_z \sin(k_z z) \\
    -i k_y k_z \sin(k_z z) \\
    |\mathbf{k}_{\parallel}|^{2} \cos(k_z z)
   \end{pmatrix}
   \hat{\alpha}_{(\mathbf{k}),1}
\quad
+ e^{-i\mathbf{k}_{\parallel}\cdot\mathbf{r}_{\parallel}}
   \begin{pmatrix}
    i k_x k_z \sin(k_z z) \\
    i k_y k_z \sin(k_z z) \\
    |\mathbf{k}_{\parallel}|^{2} \cos(k_z z)
   \end{pmatrix}
   \hat{\alpha}^{\dagger}_{(\mathbf{k}),1}
   \Biggr]+
\\
&\quad 
\frac{i|\mathbf{k}_{\parallel}|^{-1}\sin(k_z z)}{(|\mathbf{k}_{\parallel}|^{2}+ k_z^2)^{1/4}}
\begin{pmatrix}
    -k_{y} \\ k_{x} \\ 0
\end{pmatrix}
\left(
e^{i\mathbf{k}_{\parallel}\cdot\mathbf{r}_{\parallel}}\hat{\beta}_{(\mathbf{k}),2}
-
e^{-i\mathbf{k}_{\parallel}\cdot\mathbf{r}_{\parallel}}\hat{\beta}^{\dagger}_{(\mathbf{k}),2}
\right)
\Bigg\}
\end{split}
\label{eq:SI_QS_20}
\end{equation}
\noindent where \(\tilde{A}_0 = \sqrt{\frac{2\pi}{L_z}}\). The ladder operators in the continuum limit are defined as:
\begin{equation}
\alpha_{(\mathbf{k}),\lambda}
=
\sqrt{\frac{L_x L_y}{(2\pi)^2}}\,\alpha_{\mathbf{k},\lambda},
\qquad
\beta_{(\mathbf{k}),\lambda}
=
\sqrt{\frac{L_x L_y}{(2\pi)^2}}\,\beta_{\mathbf{k},\lambda}
\label{eq:SI_QS_21}
\end{equation}
\noindent leading to the continuum commutation relations:
\begin{equation}
[\alpha_{(\mathbf{k}),\lambda},\alpha_{(\mathbf{k'}),\lambda'}^\dagger]
= 
\delta(\mathbf{k}-\mathbf{k'})\,\delta_{\lambda,\lambda'}
\label{eq:SI_QS_22}
\end{equation}
\noindent Note that, by definition of $\hat{\mathbf{A}}_{\mathrm{obl}}(\mathbf{r},t)_{k_z \text{ fixed},\, k_x,k_y \neq 0}$, the points $k_{x}=0$ and 
$k_{y}=0$ are formally excluded from the integration domain. However, the integrands in equation \ref{eq:SI_QS_20} are continuous and well-behaved near $(k_{x},k_{y})=(0,0)$ and they actually vanish at the origin. Therefore, one can remove an arbitrarily small open region around $(0,0)$ without affecting the value of the integral. For this reason, we extend the domain to include the origin, which yields an equivalent and numerically convenient expression. 

\paragraph{S2.4. Definition of the photonic wavefunction}
The integrals in equation \ref{eq:SI_QS_20} are evaluated by computing the expectation value of the light-matter coupling term 
$\hat{\mathbf{p}} \cdot \hat{\mathbf{A}}$ and the diamagnetic term $\hat{\mathbf{A}}^{2}$ of the Pauli-Fierz Hamiltonian 
(Equqation \ref{eq:SI_BT_2}) over a suitable photonic state. We work in the tensor-product ansatz $\ket{\Psi_{\mathrm{pol}}} = \ket{\Psi_{\mathrm{el}}}\otimes\ket{\Psi_{\mathrm{Phot}}}$, such that electronic and photonic degrees of freedom are separable. Therefore, in the light–matter interaction term $\hat{\mathbf{p}}\cdot\hat{\mathbf{A}}$ we first evaluate the matrix element of the photonic operator $\hat{\mathbf{A}}$ over the photonic state, and only at a later stage the expectation value of the electronic momentum operator will be computed over $\ket{\Psi_{\mathrm{el}}}$. For compactness, and since the present derivation focuses on integrating out the photonic modes, the operator $\hat{\mathbf{p}}$ is not explicitly written in the intermediate steps below; its contribution factors out from the photonic expectation values due to the separability of the polaritonic wavefunction.

We consider a multimode photonic manifold where each cavity mode can host 
at most one photon (single-photon approximation). Only modes that satisfy the cavity boundary conditions are included, and the in-plane 
dependence follows the polaritonic Bloch form. 

With these assumptions, the photonic state is chosen as:
\begin{equation}
    \ket{\Psi_{\mathrm{Phot}}} =
    \Biggl(
        c_{|k_{z}|} \big( \hat{\beta}^{\dagger}_{\mathbf{0},1} + \hat{\beta}^{\dagger}_{\mathbf{0},2} \big)
        + \sqrt{1 - 2c_{|k_{z}|}^{2}}
    \Biggr)
    \Biggl(
        c_{0}
        + \sum_{\mathbf{k}_{j}}
        c_{\mathbf{k}_{j}}
        e^{-i \mathbf{k}_{||} \cdot \mathbf{r}}
        \big( \hat{\alpha}^{\dagger}_{\mathbf{k}_{j},1} + \hat{\beta}^{\dagger}_{\mathbf{k}_{j},2} \big)
    \Biggr)
    \ket{0_{\mathbf{k}_{j}},0_{\mathbf{k}_{j+1}},...,0},
    \label{eq:SI_QS_23}
\end{equation}
\noindent where the sum $\sum_{\mathbf{k}_{j}}=\sum_{\mathbf{k}_{j}=(\mathbf{k}_{\parallel},|\mathbf{k}_{z}|)}$ actually runs over the in-plane part 
$\mathbf{k}_{||} = (k_{x}, k_{y})$ of the oblique modes, being $|\mathbf{k}_{z}|$ fixed by the resonance condition of the cavity.

Thus, the photonic ansatz takes the tensor-product form:
\begin{equation}
\ket{\Psi_{\mathrm{Phot}}}
=
\ket{\Psi_{\mathbf{k_{z}}}} \otimes \Psi_{\mathrm{obl}}
\label{eq:SI_QS_24}
\end{equation}

\noindent where both components are expanded in the $\{ \ket{0}, \ket{1} \}$ Fock basis and correspond to the fundamental and oblique 
cavity-mode subspaces, respectively. The coefficients $c_{|k_{z}|}$, $\sqrt{1-2c_{|k_{z}|}^{2}}$, $c_{0}$, and $c_{\mathbf{k}_{j}}$ 
are chosen to ensure the separate normalization of $\ket{\Psi_{\mathbf{k_{z}}}}$ and $\ket{\Psi_{\mathrm{obl}}}$. More specifically, the coefficient of $\Psi_{\mathrm{obl}}$, $c_{0}$ and $c_{\mathbf{k}_{j}}$ are the square roots of the Planck probabilities of the vacuum state $\ket{0_{\mathbf{k}_{j}},0_{\mathbf{k}_{j+1}},...0}$ and the single-photon state $\ket{0,..,0_{\mathbf{k}_{j}-1},1_{\mathbf{k}_{j}},0_{\mathbf{k}_{j}+1}..,0}$. Thus, they represent the thermally weighted probabilities that all oblique modes are unoccupied $(c_{0})$ and that only the specific mode $\mathbf{k}_{j}$ is occupied while all the other remain unoccupied $(c_{\mathbf{k}_{j}})$. By recalling that the canonical partition function associated with the multimode field is $Z=\prod_{\mathbf{k}}Z_{\mathbf{k}}=\prod_{\mathbf{k}}\frac{e^{-\frac{\hbar \omega_{\mathbf{k}}}{2k_{B}T}}}{1-e^{-\frac{\hbar \omega_{\mathbf{k}}}{k_{B}T}}}$ and that the Planck factors for the vacuum state and the single-photon state are $\prod_{\mathbf{k}}e^{-\frac{\hbar \omega_{\mathbf{k}}}{2k_{B}T}}$ and $\prod_{\mathbf{k} \ne \mathbf{k'}}e^{-\frac{\hbar \omega_{\mathbf{k}}}{2k_{B}T}}e^{-\frac{\hbar \omega_{\mathbf{k'}}}{k_{B}T}}$, respectively, one obtains:

\begin{subequations}\label{eq:SI_QS_25}
\begin{align}
&c_{0}=\sqrt{P_{vac}} = \sqrt{\prod_{\mathbf{k}}(1-e^{-\frac{\hbar \omega_{\mathbf{k}}}{k_{B}T}})} \label{eq:SI_QS_25a} \\
&c_{\mathbf{k}_{j}}=\sqrt{P_{vac}}e^{-\frac{\hbar \omega_{\mathbf{k}_{j}}}{2k_{B}T}} \label{eq:SI_QS_25b}
\end{align}
\end{subequations}

So that the oblique modes state component is:

\begin{equation}
\ket{\Psi_{\mathrm{obl}}}=\sqrt{P_{vac}}\Biggl(1+\sum_{\mathbf{k}_{j}=(\mathbf{k}_{\parallel},|\mathbf{k}_{z}|)}e^{-i\mathbf{k}_{\parallel}\cdot\mathbf{r}_{\parallel}}e^{-\frac{\hbar\omega_{\mathbf{k}_{j}}}{2k_{B}T}}(\hat{\alpha}^{\dagger}_{\mathbf{k}_{j},1}+\hat{\beta}^{\dagger}_{\mathbf{k}_{j},2})\Biggr)\ket{0_{\mathbf{k}_{j}},0_{\mathbf{k}_{j+1}}..,0}
    \label{eq:SI_QS_26}
\end{equation}

\noindent which results normalized by imposing:

\begin{equation}
    P_{vac} = \frac{1}{1+2\sum_{\mathbf{k}_{j}}e^{-\frac{\hbar \omega_{\mathbf{k}_{j}}}{k_{B}T}}}
    \label{eq:SI_QS_27}
\end{equation}

\noindent To evaluate the contribution of the multimode photonic field, we proceed in two steps.  
First, we compute the expectation value of the vector potential (or its square) over the oblique-mode component
$\ket{\Psi_{\mathrm{obl}}}$.  
For both the linear term $\hat{\mathbf{p}}\cdot \hat{\mathbf{A}}$ and the diamagnetic term $\hat{\mathbf{A}}^{2}$, only oblique modes contribute in this step.  
In particular, the linear contribution involving the longitudinal mode vanishes since the longitudinal part of the vector potential does not couple with $\ket{\Psi_{\mathrm{obl}}}$:

\begin{equation}
    \langle \Psi_{\mathrm{obl}} | \hat{\mathbf{A}}_{k_z} | \Psi_{\mathrm{obl}} \rangle = 0 \rightarrow \langle \Psi_{\mathrm{obl}} | \hat{\mathbf{A}} | \Psi_{\mathrm{obl}} \rangle = \langle \Psi_{\mathrm{obl}} | \hat{\mathbf{A}}_{\mathrm{obl}} | \Psi_{\mathrm{obl}} \rangle
\label{eq:SI_QS_28}
\end{equation}

\noindent while for the quadratic term all cross terms cancel, yielding:
\begin{equation}
  \langle \Psi_{\mathrm{obl}} | \hat{\mathbf{A}}_{\mathrm{tot}}^2 | \Psi_{\mathrm{obl}} \rangle
=
\langle \Psi_{\mathrm{obl}} | \hat{\mathbf{A}}_{\mathrm{obl}}^2 | \Psi_{\mathrm{obl}} \rangle 
\label{eq:SI_QS_29}
\end{equation}

\noindent In the second step, these expectation values define an effective longitudinal operator:

\begin{equation}
\hat{\mathbf{A}}_{\mathrm{eff}}
=
\hat{\mathbf{A}}_{k_z}
+
\langle \Psi_{\mathrm{obl}} | \hat{\mathbf{A}}_{\mathrm{obl}} | \Psi_{\mathrm{obl}} \rangle     
\label{eq:SI_QS_30}
\end{equation}

\noindent and the corresponding effective Hamiltonian:

\begin{equation}
\hat{H}_{\mathrm{eff}}
=
\langle \Psi_{\mathrm{obl}} | \hat{H} | \Psi_{\mathrm{obl}} \rangle 
\end{equation}

\noindent The expectation value is then evaluated on $\ket{\Psi_{k_z}}$.  
The coefficients of $\ket{\Psi_{k_z}}$, in particular $c_{|k_z|}$, are treated as variational parameters and are optimized by minimizing the reduced energy:

\begin{equation}
    E_{\mathrm{eff}}(c_{|k_z|})=
\langle \Psi_{k_z} |\, \hat{H}_{\mathrm{eff}}\, | \Psi_{k_z} \rangle
\label{eq:SI_QS_31}
\end{equation}

\noindent subject to the normalization constraint.  
This procedure ensures that the longitudinal cavity mode is optimally dressed by the continuum of oblique modes, after integrating out their degrees of freedom, providing a self-consistent multimode vacuum renormalization.

\paragraph{S2.5. Evaluation of the contribution of the multimode field on the energy}

Under the single-photon approximation, the only non-null terms associated to $\langle \Psi_{\mathrm{obl}} | \hat{\mathbf{A}}_{\mathrm{obl}} | \Psi_{\mathrm{obl}} \rangle $ are $\bra{0} \hat{\alpha}_{\mathbf{k},1}\{\hat{\beta}_{\mathbf{k},2}\} \ket{1_{\mathbf{k},1/2}}$ and $\bra{1_{\mathbf{k},1/2}} \hat{\alpha}^{\dagger}_{\mathbf{k},1}\{\hat{\beta}^{\dagger}_{\mathbf{k},2}\} \ket{0}$, while those associated to  $\langle \Psi_{\mathrm{obl}} | \hat{\mathbf{A}}_{\mathrm{obl}}^{2} | \Psi_{\mathrm{obl}} \rangle $ are 
$\bra{1_{\mathbf{k},1/2}}{\hat{\alpha}^{\dagger}_{\mathbf{k},1}\hat{\alpha}_{\mathbf{k},1}\{\hat{\beta}^{\dagger}_{\mathbf{k},2}\hat{\beta}_{\mathbf{k},2}\}}\ket{1_{\mathbf{k},1/2}}$ and $\bra{0}{\hat{\alpha}_{\mathbf{k},1}\hat{\alpha}^{\dagger}_{\mathbf{k},1}\{\hat{\beta}_{\mathbf{k},2}\hat{\beta}^{\dagger}_{\mathbf{k},2}\}}\ket{0}$. Therefore, in polar coordinates, the expectation values $\langle \Psi_{\mathrm{obl}} | \hat{\mathbf{p}} \cdot\hat{\mathbf{A}}_{\mathrm{obl}} | \Psi_{\mathrm{obl}} \rangle $ and $\langle \Psi_{\mathrm{obl}} | \hat{\mathbf{A}}_{\mathrm{obl}}^{2} | \Psi_{\mathrm{obl}} \rangle $ read:

\begin{equation}
  \langle \Psi_{\mathrm{obl}} | \hat{p} \cdot \hat{\mathbf{A}}_{\mathrm{obl}} | \Psi_{\mathrm{obl}} \rangle= 4\pi\Tilde{A}_{0}\cos({k_{z}z})p_{z}P_{vac}\int_{0}^{+\infty}dk_{\parallel}\frac{k^{2}_{\parallel}e^{-\frac{ c}{2T}
\sqrt{k^{2}_{\parallel}+k^{2}_{z}}}}{\sqrt[4]{(k^{2}_{\parallel}+k^{2}_{z}})^3}  
\label{eq:SI_QS_32}
\end{equation}

\begin{equation}
\begin{split}
    \langle \Psi_{\mathrm{obl}} | \hat{\mathbf{A}}_{\mathrm{obl}}^2 | \Psi_{\mathrm{obl}} \rangle =& 8\pi\Tilde{A}^{2}_{0}P_{vac}\int_{0}^{+\infty}dk_{\parallel}\frac{k^{2}_{z}\sin^2(k_{z}z)k_{\parallel}
    +\cos^2(k_{z}z)k^{3}_{\parallel}}{\sqrt{(k^{2}_{\parallel}+k^2_{z})^3}}e^{-\frac{c}{T}\sqrt{k^{2}_{\parallel}+k^2_{z}}} + \\
    &8\pi\Tilde{A}^{2}_{0}P_{vac}\int_{0}^{+\infty}dk_{\parallel}\frac{\sin^2(k_{z}z)k_{\parallel}
    }{\sqrt{k^{2}_{\parallel}+k^2_{z}}}e^{-\frac{c}{T}\sqrt{k^{2}_{\parallel}+k^2_{z}}} + \\
    &2\pi\Tilde{A}^{2}_{0}P_{vac}\int_{0}^{+\infty}dk_{\parallel}\frac{k^{2}_{z}\sin^2(k_{z}z)k_{\parallel}
    +\cos^2(k_{z}z)k^{3}_{\parallel}}{\sqrt{(k^{2}_{\parallel}+k^2_{z})^3}} +  \\&2\pi\Tilde{A}^{2}_{0}P_{vac}\int_{0}^{+\infty}dk_{\parallel}\frac{\sin^2(k_{z}z)k_{\parallel}
    }{\sqrt{k^{2}_{\parallel}+k^2_{z}}}
\end{split}
\label{eq:SI_QS_33}
\end{equation}

\noindent All the integrands are regular and integrable, with the exception of the last two terms of equation \ref{eq:SI_QS_33}, namely
$2\pi\Tilde{A}^{2}_{0}P_{vac}\int_{0}^{+\infty}dk_{\parallel}\,\frac{\cos^{2}(k_{z}z)\,k_{\parallel}^{3}}{\sqrt{(k_{\parallel}^{2}+k_{z}^{2})^{3}}}$
and
$2\pi\Tilde{A}^{2}_{0}P_{vac}\int_{0}^{+\infty}dk_{\parallel}\,\frac{\sin^{2}(k_{z}z)\,k_{\parallel}}{\sqrt{k_{\parallel}^{2}+k_{z}^{2}}}$.
In fact, these two contributions exhibit ultraviolet divergence for $k_{\parallel}\rightarrow \infty$ and therefore require regularization.

However, both integrals diverge as $\sqrt{k_{\parallel}^{2}+k_{z}^{2}}$.  
Thus, even without applying the LWA, the diamagnetic term becomes independent of $z$, since the same divergent factor multiplies $\sin^{2}(k_{z}z)$ and $\cos^{2}(k_{z}z)$ and their sum removes any spatial dependence.

To regularize this divergence, we introduce a cutoff at $\sqrt{k_{\parallel}^{2}+k_{z}^{2}}/c=\omega_{p}$, the plasma frequency of the metallic mirrors. Modes above $\omega_{p}$ are not confined by the cavity and therefore do not contribute to the vacuum field, making this cutoff physically well-justified.

Concerning instead the linear term $ \langle \Psi_{\mathrm{obl}} | \hat{p} \cdot \hat{\mathbf{A}}_{\mathrm{obl}} | \Psi_{\mathrm{obl}} \rangle$, it is calculated by numerical integration (see materials and methods M2).

\end{document}